\documentclass[twocolumn,fleqn]{autart}    

\usepackage{graphicx}          

\usepackage{amsmath}
\usepackage{color,subfig}
\let\dfrac\frac

\newcommand{\ii}{\mathrm{i}}
\newcommand{\Matrix}[1]{\left[\begin{matrix}#1\end{matrix}\right]}
\newcommand{\hide}[1]{}

\begin{document}

\begin{frontmatter}

\title{Accurate Frequency Domain Identification of ODEs with Arbitrary Signals}



\thanks[footnoteinfo]{Corresponding author Eduardo.~Martini. Tel.  +33549366009. 
Fax +33549366000.}

\author[ita,pprime]{Eduardo Martini}\ead{emartini@ita.br},
\author[ita]{Andr\'e V. G. Cavalieri}\ead{andre@ita.br},
\author[pprime]{Peter Jordan}\ead{peter.jordan@univ-poitiers.fr},
\author[poly]{Lutz Lesshafft}\ead{lutz@ladhyx.polytechnique.fr}

\address[ita]{Instituto Tecnol\' ogico de Aeron\'{a}utica,   S\~ ao Jos\' e dos Campos/SP, Brazil}
\address[pprime]{ D\'epartement Fluides, Thermique et Combustion, Institut Pprime, CNRS, Universit\' e de Poitiers,ENSMA, 86000 Poitiers, France }
\address[poly]{Laboratoire d'Hydrodynamique, CNRS, Ecole Polytechnique, Institut Polytechnique de Paris, 91128 Palaiseau, France}

\begin{keyword}                           
Identification methods, Time-invariant, System identification
\end{keyword}                             

\begin{abstract}                          
The difficulty in frequency domain identification is that frequency components of arbitrary inputs and outputs are not related by the system’s transfer function if signals are windowed. When rectangular windows are used, it is well known that this difference is related to transient effects that can be estimated alongside the systems’ parameters windows. In this work, we generalize the approach for arbitrary windows, showing that signal windowing introduces additional terms in the system’s equations. The formalism is useful for frequency-domain input-output analysis of a system, and also for system identification. For the latter application, the approach considerably reduces aliasing effects and allows the computation of the associated correction terms, reducing the number of parameters that need to be estimated. The system identification approach has features of the modulating-function technique, filtering out the effects of initial conditions while retaining the spectral interpretation of frequency-domain methods. 
\end{abstract}

\end{frontmatter}

\section{Introduction}

\label{sec:introduction}
Most physical systems are mathematically described by systems of differential equations, whose dynamics can be obtained or modelled by identification techniques: identification of the discretized system, frequency-domain identification, modulating functions, among others. Textbook approaches for frequency-domain identification involve the application of periodic inputs, allowing for the application of fast and accurate techniques \cite{pintelon2012system}. The use of arbitrary signals requires spurious transients to be accounted for \cite{pintelon1997frequency,pintelon1997identification}. 

Here we focus on fully observable linear systems described by ordinary differential equations, as
\begin{equation}\label{eq:sysDiffEq}
\sum_{j=0}^{n_a} A_j \frac{d^jx}{dt^{j}} = \sum_{k=0}^{n_b} B_k \frac{d^k u}{dt^{k}} ,
\end{equation}
where $ x $ is the state vector with size $ n_x $, $ u $ the input vector with size $ n_u $, $ A_j $, and $ B_k $ are system matrices with sizes $ n_x\times n_x  $ and $ n_x\times n_u $, with $ n_u\le n_x $. Inputs and responses are sampled  with  $ N $ equally spaced points on a time interval between 0 and T,  $ t_j = jT/N $, and corresponding sampling,  $ f_s=N/T $, and Nyquist, $ f_{nyq}=N/(2T) $, frequencies.	A frequency-domain representation of \eqref{eq:sysDiffEq} reads
\begin{equation}\label{eq:sysDiffEqfreq}
	L(f) \hat x(f) = R(f) \hat u(f) + \hat e(f),
\end{equation}
where 
\begin{align} \label{eq:fouTrans}
	\hat x(f) =& \int_{-\infty}^\infty x(t)\mathrm{e}^{-2\pi \mathrm{i} f t}dt , \\
	\hat u(f) =& \int_{-\infty}^\infty u(t)\mathrm{e}^{-2\pi \mathrm{i} f t}dt , \\ 
	L(f) =& \sum_{j=0}^{n_a} (-2\pi \ii f)^j A_j,   \\
	R(f) =& \sum_{k=0}^{n_b} (-2\pi \ii f)^k B_k.
\end{align}
An error term, $\hat e(f)$, was introduced to account for errors due to signal noise, windowing, and finite sampling rates, which will be discussed later.

Time-domain identification consists of estimating matrices $ A_j $ and $ B_j $ from $ x(t) $ and $ u(t) $ data, while frequency domain identification approaches the problem via their spectral components, $ \hat x(f) $ and $\hat u(f) $. These approaches, although equivalent in theory, have significant practical differences. For instance, coloured (non-white) time-invariant noise generates signals that are correlated in time, and optimal time-domain identification requires the use of full correlation matrices. In the frequency domain the components are uncoupled, which has important practical advantages \cite{mckelvey2000frequency}. If periodic signals can be used, measurements of the system transfer function can be easily performed and used for different system-identification methods, e.g.  \cite{mi2012frequencydomain,soderstrom2017errorsinvariables}.

However, when non-periodic signals are used with finite data lengths, spectral components of the inputs and outputs lead to non-zero errors in \eqref{eq:sysDiffEqfreq}. Such error have been classically associated with spectral leakage, although later \cite{pintelon1997frequency} and \cite{pintelon1997identification} showed they can be understood as spurious transient effects. It was shown that, for rectangular windows, the error is a polynomial term of order $ n_p=\max(n_a,n_b) $. The system parameters and polynomial coefficients can be estimated simultaneously, allowing for accurate identifications to be obtained. 
In continuous-time models, finite sampling rates lead to aliasing errors, which were minimized in \cite{pintelon1997identification} by artificially increasing the polynomial order of the correction term.  Reference \cite{schoukens2006leakage} shows that systematic plant estimation errors scale with $ 1/N $ when rectangular windows are used, or an improved convergence of $ 1/N^2 $ when Hanning or \textit{Diff} windows are used. System identification can be performed as proposed by \cite{pintelon1997frequency,pintelon1997identification}, or via variations of the method, e.g.  \cite{vanherpen2014optimally,soderstrom2017errorsinvariables}.

Focusing on continuous systems, a different approach consists in multiplying \eqref{eq:sysDiffEq} by modulating functions and integrating over time. By choosing modulating functions that have their first $ n $-th derivatives equal to zero at their limits, where $ n=\max(n_a,n_b) $ is the order of the system \eqref{eq:sysDiffEq}, integration by parts eliminates effects of initial conditions and also avoid the need to compute derivatives of the system's input and output. The terms  $ A_j $ and $ B_k $ can be estimated from the resulting linear system obtained using several such modulating functions.  Various modulating functions have been used, such as spline \cite{preisig1993theory}, sinusoidal \cite{co1990system}, Hermite polynomials \cite{takaya1968use}, wavelets \cite{sadabadi2008system}, and Poisson moments \cite{saha1982structure}. Modulating functions have been used in the identification of integer and fractional-order systems \cite{nazarian2010identifiability} and extended to identify both model parameters and model inputs from response observations only \cite{asiri2017modulating}.

We propose a different interpretation of transient effects in the frequency domain. Correction terms for the time derivatives to compensate for windowing effects are derived, considerably reducing the error term in \eqref{eq:sysDiffEqfreq}, with the remaining errors being due to aliasing effects and signal noise.  The  corrections can be understood as spurious inputs, which reproduces the polynomial term  introduced by \cite{pintelon1997identification}  when rectangular windows are used.  The proposed method however allows for the use of arbitrary windows, with spurious terms due to signal windowing being computed instead of estimated. This also drastically reduces aliasing effects, avoiding the need to account for those with artificial terms.

The paper is structured as follows.  In section \ref{sec:sigwindowing} effects of signal windowing on ODEs are analysed and correction terms are derived. The approach is explored in section \ref{sec:sysIden} for purposes of system identification. A classification of two types of aliasing effects is proposed in section \ref{sec:aliasing}, with two classes of windows that minimize one type of aliasing are presented in section \ref{sec:window}, being  one of those a novel infinity smooth window. Numerical experiments are presented in section \ref{sec:numTests}. Conclusions are presented in section \ref{sec:conclusion}. Details on the source of aliasing errors are presented in appendix \ref{app:aliasing}.

\section{Effects of signal windowing and system identification}
\subsection{Signal windowing} \label{sec:sigwindowing}
In practice,  $ \hat x(f)  $ and $ \hat{u}(f) $, \eqref{eq:fouTrans}, are estimated from windowed signals as, 
\begin{align}\label{}
	\label{eq:fourrier_overlinex} \hat x_w(f) =& \int_0^T w(t)x(t)\mathrm{e}^{-2\pi \mathrm{i} f t}dt ,\\
	\label{eq:fourrier_overlineu} \hat u_w(f) =& \int_0^T w(t)u(t)\mathrm{e}^{-2\pi \mathrm{i} f t}dt,
\end{align}
where $ w(t) $ is a window function. For frequencies $ f = j/T $, $ \hat{x}_w(f) $  and $ \hat{u}_w(f) $ coincide with Fourier-series coefficients of the periodic extension of $ (wx) $ and $ (wu) $. These values are typically obtained by performing a fast Fourier transform (FFT) on discrete-time samples. 

Even in the absence of aliasing effects and noise, using $\hat x_w$ and $\hat u_w$ in \eqref{eq:sysDiffEqfreq} leads non-zero errors. Multiplying  \eqref{eq:sysDiffEq} by the window function $w(t)$,
\begin{equation}\label{eq:sysDiffEq_w}
	\sum_{j=0}^{n_a} A_j \left( w(t) \dfrac{d^jx}{dt^{j}} \right) = \sum_{k=0}^{n_b} B_k \left( w(t) \dfrac{d^k u}{dt^{k}} \right),
\end{equation}
and defining $	{x}^{\{0\}}(t) =0$,
\begin{align}\label{eq:correctionTerm}
	{x}^{\{j\}}(t) = & \dfrac{d^j(wx)}{dt^{j}}(t) - w(t)\left(\dfrac{d^j {x}}{dt^j}\right)(t),
\end{align}                                               
for $j>0$, and analogous expressions for $u(t)$,
\eqref{eq:sysDiffEq_w} can be re-written as 
\begin{equation}
	\sum_{j=0}^{n_a} A_j \left(  \dfrac{d^j (wx) }{dt^{j}} - {x}^{\{j\}} \right) = \sum_{k=0}^{n_b} B_k \left(  \dfrac{d^k (wu)}{dt^{k}}  - {u}^{\{k\}} \right).
\end{equation}
Applying a Fourier transform of \eqref{eq:sysDiffEq_w2} leads to
\begin{equation}\label{eq:sysDiffEq_w2}
	\sum_{j=0}^{n_a} A_j \left(  (-2\pi\ii)^j  \hat x_w  - \hat {x}^{\{j\}} \right) = \sum_{k=0}^{n_b} B_k \left(  (-2\pi\ii)^k \hat u_w - \hat {u}^{\{k\}} \right),
\end{equation}
where the frequency dependence was omitted for clarity. 
Comparing to \eqref{eq:sysDiffEqfreq} it can be seen that the terms $\hat {x}^{\{j\}} $ and $\hat {u}^{\{j\}} $ are corrections that appear when Fourier transforms of windowed signals are used instead of the true Fourier transforms of the signals.  Re-arranging \eqref{eq:sysDiffEq_w2} as 
\begin{equation}\label{eq:sysDiffEq_spurrious}
	L \hat x_w= R\hat u_w + \left(\sum_{j=0}^{n_b} A_j \hat {x}^{\{j\}}  - \sum_{k=0}^{n_b} B_k \hat {u}^{\{k\}} \right) .
\end{equation}
The term in parenthesis, which will be referred to as \emph{spurious inputs}, constitutes a significant contribution to the error in \eqref{eq:sysDiffEqfreq} if not accounted for. Note also that these errors are correlated with the signals of $u$ and $x$, and thus not accounting for them introduces a significant bias in system estimation \cite{pintelon1997frequency,pintelon1997identification}. Signal noise and finite sampling contribute to error terms in  \eqref{eq:sysDiffEq_spurrious} and \eqref{eq:sysDiffEq_w2}, similar to those of \eqref{eq:sysDiffEqfreq}.

We describe a procedure to compute $  x^{\{j\}}  $, with the computation of $  u^{\{k\}}  $ being analogous.  It is useful to express  $  x^{\{j\}}  $ as a sum of terms of the form, $ \frac{d^m}{dt^m}(\frac {d^kw}{dt^k} x) $, as the window derivative can be obtained analytically and the outer derivative obtained in the frequency domain,  avoiding the computation of time derivatives of $ x $. A recurrence relation for $ x^{\{j\}}$ is obtained by noting that,
 \begin{equation}\label{eq:corrTerms}
 	\begin{aligned}
 		\dfrac{d^j x^{\{i\}} } {dt^j}  &= \dfrac{d^{i+j} (wx) }{dt^{i+j}} - \dfrac{d^j(w d^ix/dt^i)}{dt^j} \\
 		& =\sum_{k=1}^{i+j} \left(\binom{k}{i+j}-\binom{k}{j} \right)  \dfrac{d^kw}{dt^k}\dfrac{d^{i+j-k}x}{dt^{i+j-k}},
 	\end{aligned}
 \end{equation}
 where $ \binom{i}{j} $ is the binomial of $ i$ and $ j $, with the convention that $ \binom{i}{j}=0 $ for $ i<0 $ or $ i>j $. Solving
 \begin{align}\label{kex}
 	\sum_{j=0}^{n-1} a_j \dfrac{d^jx^{\{n-j\}}}{dt^j} &= \dfrac{d^nw^{n}}{dt^n} x,
 \end{align}
 as a linear system 
 \begin{align}\label{kex}
 	\sum_ {j=0}^{n-1} A_{i,j} a_j  = & \delta_{i,n}, \\
 	A_{i,j} = & \binom{i}{n}-\binom{i}{j} .
 \end{align}
 allows $ x^{\{i\}} $ to be obtained as
 \begin{equation}\label{kex}
 	x^{\{i\}} = \dfrac{1}{a_{0}} \left(\dfrac{d^i w}{dt^i} x +\sum_{j=1}^{i-1} a_j \dfrac{d^{j}x^{\{i-j\}}}{dt^j}  \right).
 \end{equation}
 
 The first three correction terms read
 \begin{align}\label{kex}
 	x^{\{1\}} &= \;\;\;\dfrac{dw}{dt}x   ,
 	\\x^{\{2\}} &=   -   \dfrac{d^2w}{dt^2}x +2 \dfrac{dx^{\{1\}}}{dt}  ,
 	\\x^{\{3\}} &= \;\;\;\dfrac{d^3w}{dt^3}x +3 \dfrac{dx^{\{2\}}}{dt} - 3 \dfrac{d^2x^{\{1\}}}{dt^2}   ,
 \end{align}
 with corresponding frequency counterparts,
 \begin{align}\label{kex}
 	\tilde x^{\{1\}} &= \;\;\;  \mathcal F\left(\dfrac{dw}{dt}x    \right),
 	\\\tilde x^{\{2\}} &=   -   \mathcal F\left(\dfrac{d^2w}{dt^2}x\right) +2 (-2\pi\mathrm{i}f)\tilde x^{\{1\}} ,
 	\\\tilde x^{\{3\}} &= \;\;\; \mathcal F\left(\dfrac{d^3w}{dt^3}x\right) +3 (-2\pi\mathrm{i}f)\tilde x^{\{2\}} - 3(-2\pi\mathrm{i}f)^2  \tilde x^{\{1\}}  , 
 \end{align}                                                                                                 
 where $ \mathcal F $  represents  Fourier transforms as in \eqref{eq:fouTrans}. 
 
 A direct connection between the derivation above and that of \cite{pintelon1997identification} is made by writing a rectangular window as a sum of Heaviside step functions, $H$, 
 \begin{align}
 	w_{rect}(t)=H(t) - H(t-T),
 \end{align}
where $T$ is the window length. The correction terms are a function of the window derivatives, which for the Heaviside step function are the delta distribution and its derivatives. The correction terms are thus polynomials whose coefficients are a function of the signal and it derivatives at 0 and T. As obtaining these coefficients directly from the data can lead to large errors they were instead  estimated a posteriori  \cite{pintelon1997frequency,pintelon1997identification}. 

With the proposed approach, we generalize the results of \cite{pintelon1997frequency,pintelon1997identification} to arbitrary windows, with the use of smooth windows allowing for the computation of the correction terms from data, reducing the number of parameters that need to be estimated.
%

\subsection{Use for system identification} \label{sec:sysIden}

Interpreting the correction terms as a correction for the time derivatives on windowed signals,  \eqref{eq:sysDiffEq_w2} can be explored to obtain an estimation of the system's parameters.
As the terms  $\hat x_w $, $\hat x^{\{i\}} $, $\hat u_w $, and $\hat u^{\{j\}} $ can be computed directly from the inputs and outputs, the system parameters $A_i$ and $B_j$ can be estimated from them. Defining the matrices
\begin{align}
\theta &= [A_0,\dots,A_{n_a},B_0,\dots,B_{n_b}],	\\
	 M &= 
	\left[ \begin{matrix}		
	\;	\mathcal{L}_{n_a}(f_0) &\dots& \;\mathcal{L}_{n_a}(f_n) \\
		\vdots& &\vdots\\
		\;\mathcal{L}_0(f_0) &\dots& \;\mathcal{L}_0(f_n) \\
		-\mathcal{R}_{n_b}(f_0) &\dots& -\mathcal{R}_{n_b}(f_0) \\
		\vdots& &\vdots\\
		-\mathcal{R}_{0}(f_0) &\dots& -\mathcal{R}_{0}(f_n) 
	\end{matrix}
	\right],		
\end{align}
where
\begin{align}
	\mathcal{L}_i(f) =& (-2\pi\omega f)^i	\hat{x}_w(f) -\hat{x}^{\{i\}}(f), \\
	\mathcal{R}_i(j) =&	(-2\pi\omega f)^i\hat{u}_w(f) -\hat{u}^{\{i\}}(f),
\end{align}
ignoring the error terms,  \eqref{eq:sysDiffEq_w2} is re-written as
\begin{align}
	 \theta M =& 0,
\end{align}
To solve for $\theta$, we fix $A_{n_a}=I$, and write 
\begin{align}
	M &= \Matrix{ M_1 \\ M_2} ,\\ 	
	\theta &= \Matrix{ \theta_1 & \theta_2} . 
\end{align}
where the terms with subscript 1 contains the first $n_x$ lines of the $M$ and the first $n_x$ columns of $\theta$. As ${M_1=A_{n_a}}$, the unknown system's parameters are all contained in $M_2$, and satisfy
\begin{align} \label{eq:M2eq}
		  \theta M_2 =& -M_1.
\end{align}
An estimation of $\theta_2$, $\tilde \theta_2$, is obtained as a least square-error solution of \eqref{eq:M2eq},  as
\begin{align}
	\tilde \theta =& -M_1 M_2^+ ,
\end{align}
 where the superscript $+$ reefers to the Moore-Penrose pseudo-inverse. 
 
 If signal noise is assumed, \eqref{eq:M2eq} becomes an error-in-variables problem, where only $ \overline M_2 $ and $ \overline M_1 $ are observable, such that
\begin{align} \label{eq:eerinvar}
	 \overline M_2 =& M_2 + M_2' , 	\\
	 \overline M_1 =& M_1 + M_1' ,
\end{align}
with $M_{1,2}'$ being the contribution to signal noise in $M_{1,2}' $. In this case, the linear-regression estimation used here is biased. The bias can be removed by several error-in-variables estimation methods proposed in the literature, e.g. \cite{soderstrom2011generalized,soderstrom2017errorsinvariables}. As the signal windowing introduces significant correlation between different frequency-domain components of the error, the resulting non-linear problems that can provide an unbiased system-estimation is more complex than the one solved in \cite{pintelon1997frequency,pintelon1997identification}, which is a trade-off for the significantly lower number of parameters that need to be estimated. In this work, we instead focus on estimations using a least-square errors approach, which, although biased, can be still used to obtain estimations with lower computational cost, and with high accuracy for low-noise scenarios.
 Note also that if the sampling frequency is low $M_{1,2}' $, will also contain aliasing effects, which, due to spectral leakage, are correlated with $M_{1,2}$: this effect can however be mitigated with the use of low-pass filters.

\section{Minimizing aliasing effects} 
\subsection{Magnitude of the aliasing effects}\label{sec:aliasing}
Equations \eqref{eq:sysDiffEq_w2} and \eqref{eq:sysDiffEq_spurrious} are exact for noiseless signals in the continuous time domain. In practice however the Fourier integrals have to be computed from sampled data, with aliasing effects leading to errors in the estimation of the Fourier-series coefficients, and consequently errors in \eqref{eq:sysDiffEq_w2} and \eqref{eq:sysDiffEq_spurrious}.  We proceed analysing the errors when estimating the Fourier coefficients of $s(t)$, which can represent $x(t)$ or $u(t)$ windowed by $d^i w/dt^i$.

To analyse the behaviour of the errors for large number of samples $N$, the convergence of the truncated Fourier representation and of the Fourier coefficients are used. Assuming $s(t)\in C^n$, the errors of truncated Fourier representation, $s_N(t)$ of $s(t$) , follow
\begin{align}
	|| s(t) - s_N(t) || < O(1/N^n),
\end{align}
for large $N$, where  $||\cdot||$ represents the standard $L_2$ norm.  The errors in the Fourier-series coefficient, $a_k$, of the periodic extension of $s(t)$, when estimated using $N$ points at finite sampling rate, is given by \cite{trefethen2014exponentially},
\begin{align} 
	|a _{k}- \tilde{a} _{k,N}  | &= O(1/N^n).
\end{align}
Note that for finite  $k$, the above expressions imply a algebraic decays, while if $s(t)\in C^\infty$, a non-algebraic decay is obtained.
Note also that if $s(t)$ is smooth everywhere with the exception of the window limits, computing the Fourier coefficients of its periodic extension with the FFT method is equivalent to an integration using the trapezoidal rule. It can then be shown that  $|a _{k}- \tilde{a} _{k,N}  | = O(1/N^{n+1}) $ for odd $n$ \cite{trefethen2014exponentially}.

As typically inputs and outputs of the system are smooth functions, the asymptotic errors in \eqref{eq:sysDiffEqfreq} due to finite sampling are given by the smoothness of the window functions used. Note that if $w(t)\in C^n$, then $d^i w/dt^i(t)\in C^{n-i}$. This means that the errors in \eqref{eq:sysDiffEqfreq} due to finite sampling decays at least with $1/N^{n-max(n_a,n_b)}$.
Estimates for $ |a _{k}- \tilde{a} _{k,N}  |  $ are derived in appendix  \ref{app:aliasing}, were it is explicitly seen that they are related to aliasing effects.

As signal windowing causes spectral leakage, which may leak spectral content in frequencies above the Nyquist frequency,
it is useful to distinguish between two types of aliasing effects: \emph{type I}, due to the  signal content at frequencies higher than the Nyquist frequency; and \emph{type II}, due to spectral leakage of the signal above the Nyquist frequency due to windowing.

Type I aliasing effects can be easily reduced with the use of spatial filters. The use of filters in $u$ and $x$ does not affect the structure of \eqref{eq:sysDiffEqfreq}, and thus does not affect frequency domain analysis or the identification of the system parameters. Filters can easily provide very fast decay of the spectra  $\hat x$, and thus the decay of $\hat w$ is the dominant factor in the type II aliasing effects. In the following subsection we propose families of windowing functions with convenient properties for the present frequency-domain analysis.

\subsection{ Windows for algebraic and non-algebraic decay of aliasing effects}\label{sec:window}
\begin{figure}
	\centering
	\includegraphics[width=\linewidth,trim={0 2px 0 2px },clip]{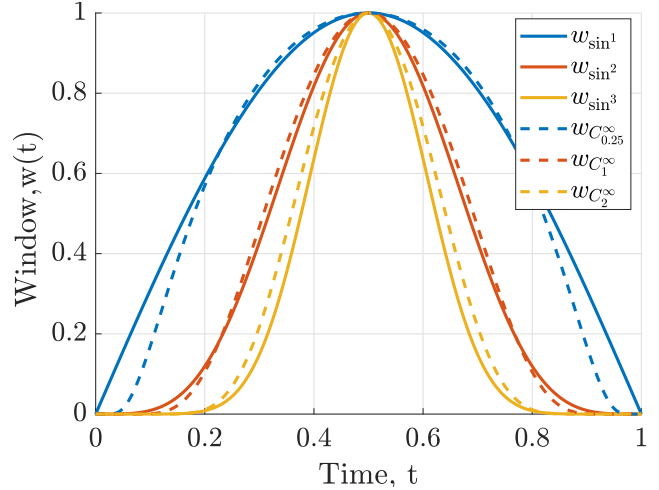}
	\caption{Proposed windows: $ w_{sin^n}(t)  $ and $ w_{C^\infty_n}(t)  $ for $ T=1 $.}
	\label{fig:windows}
\end{figure}

\begin{figure*}
	\centering
	\subfloat[ Spectral content of $w_{\sin^n}$. ]{ \includegraphics[width=.45\linewidth]{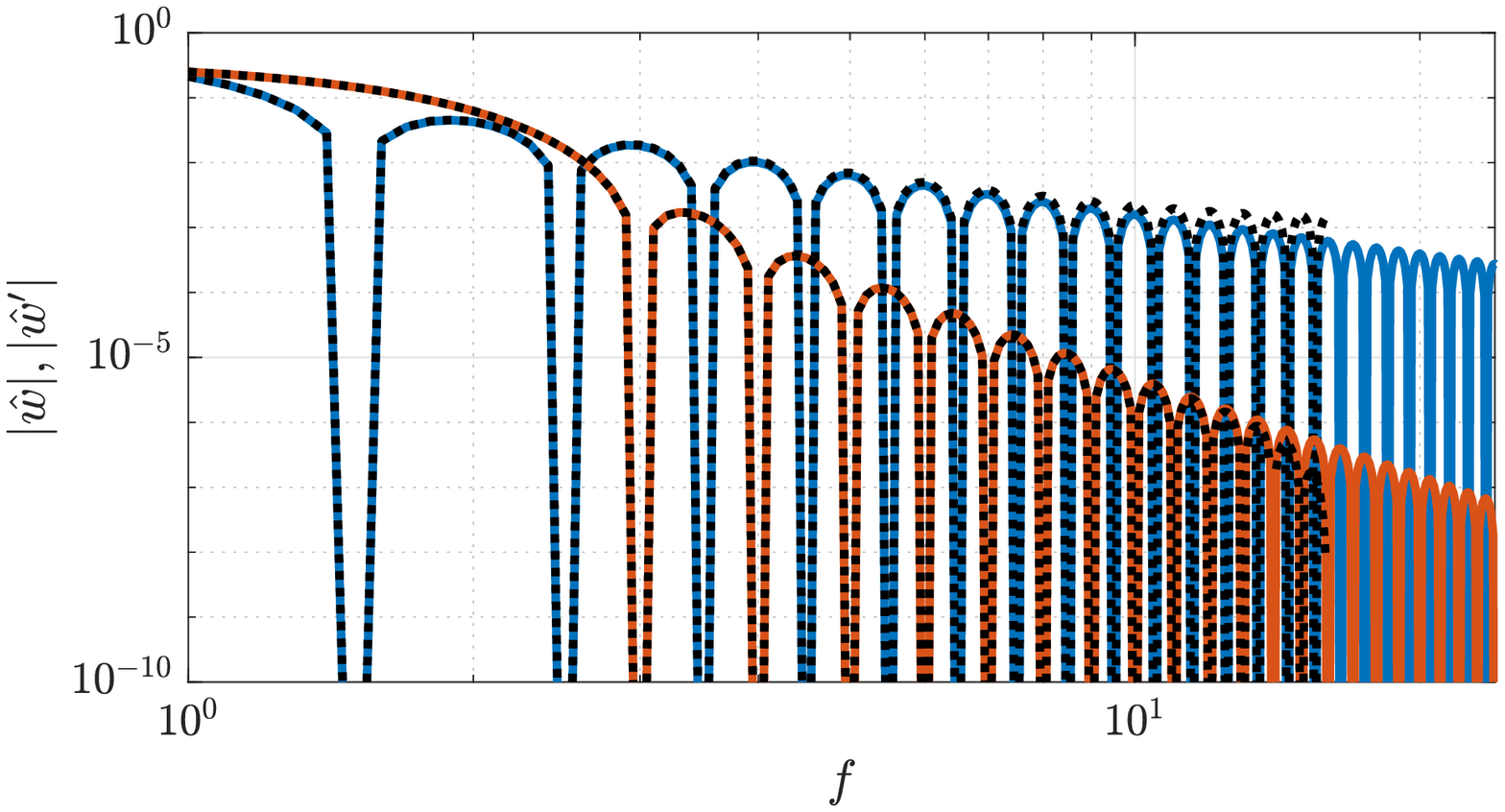} }
	\subfloat[ Spectral content of $w_{C^\infty_n}$. ]{ \includegraphics[width=.45\linewidth]{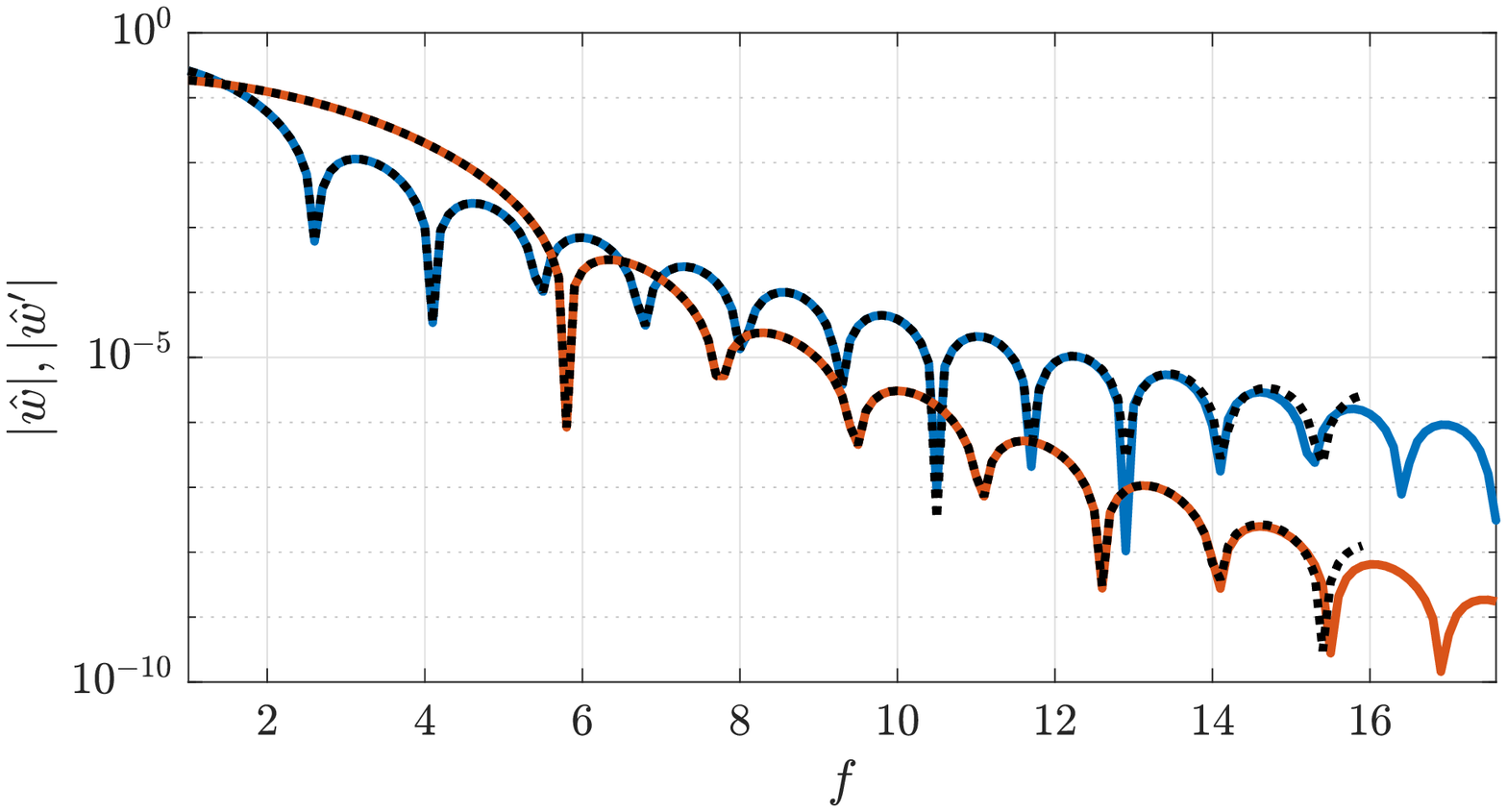} }
	
	\subfloat[ Differences for $w_{sin^n}$.]{ \includegraphics[width=.45\linewidth]{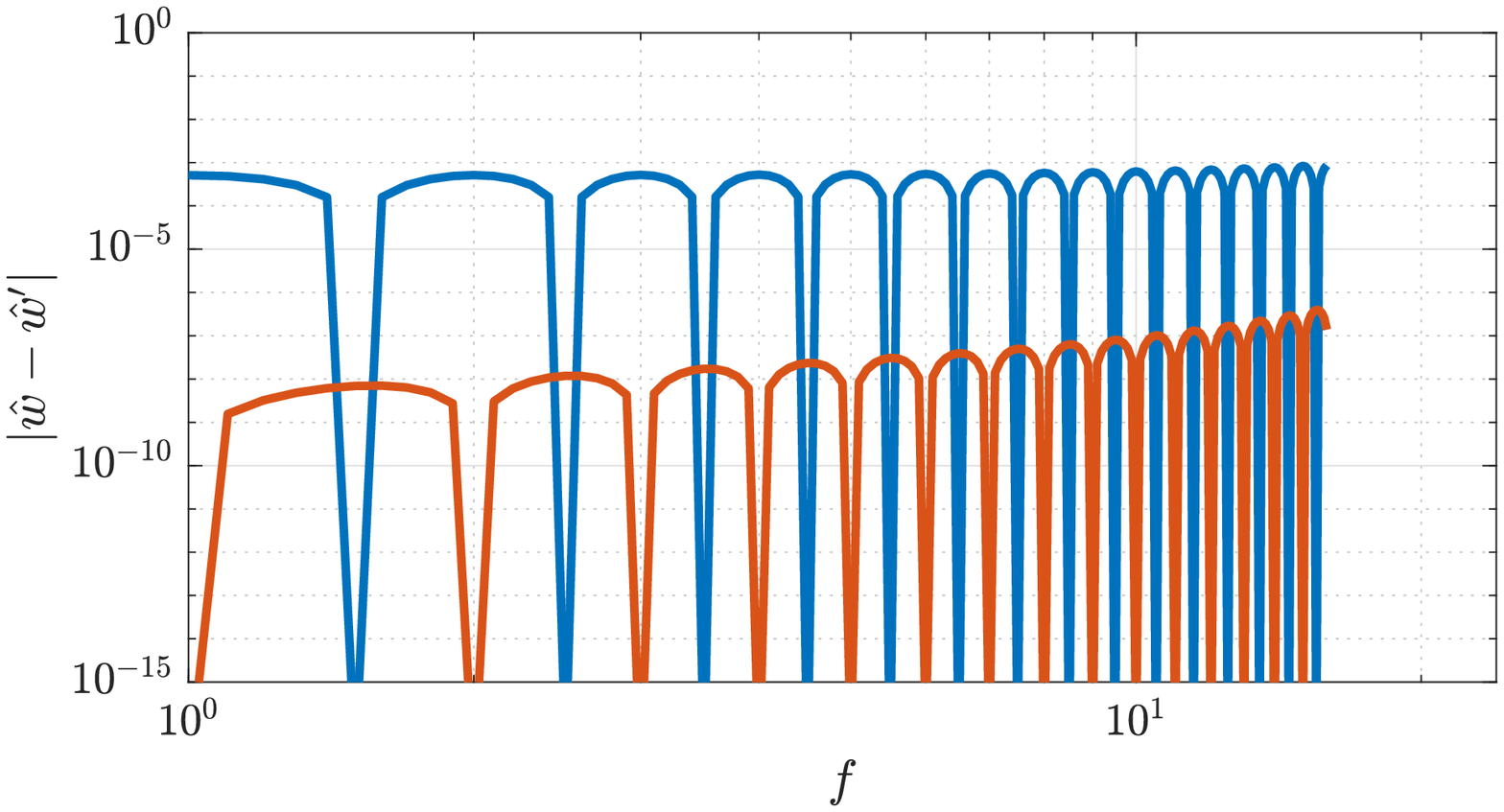} }
	\subfloat[ Differences for $w_{C^\infty_n}$. ]{ \includegraphics[width=.45\linewidth]{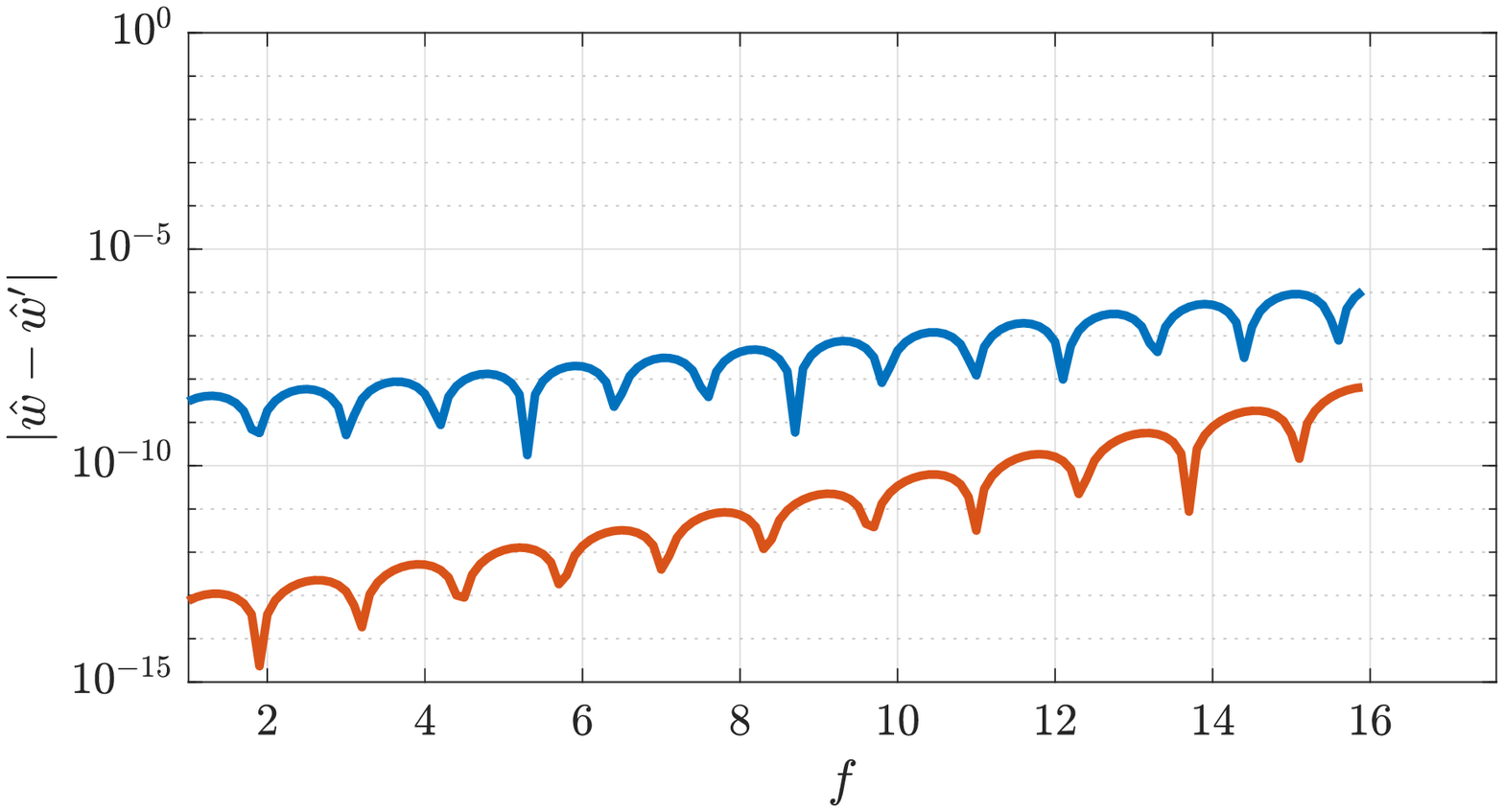} }
	
	\caption{Spectral content of the proposed windows (top) with sampling frequencies of 1024 ($\hat w$, coloured lines) and 32 ($\hat w'$, black dots). The difference between results from the two sampling frequencies are shown on the bottom figures. Blue and red lines are curves for $n=1$ and $4$, respectively.  }
	\label{fig:winspec}
\end{figure*}

Since the decay rate of the magnitude of the Fourier coefficients of the window is directly related to its smoothness \cite{schramm1982magnitude}, motivating the investigation of two window families, 
\begin{equation}\label{eq:w_sin}
	w_{\sin^n}(t) = \begin{cases} 
								\sin^n(\pi t/T) &, 0<t<T, \\
											0 &, \text{otherwise}
													\end{cases}
\end{equation}
which is $C^n$, with its first $n-1$ derivatives equal to zero at $ 0 $ and $ T $, and corresponding to the $ cos^n $ windows in \cite{harris1978use}, and a  novel infinitely-smooth window given by
\begin{equation}\label{eq:w_c}
	w_{C^\infty_n}(t) =
	 \begin{cases} 
		 {\mathrm{e}^{-\dfrac{nT^2}{t(T-t)}} }/ {\mathrm{e}^{-4 n}}, &, 0<t<T, \\
		0 &, \text{otherwise}
	\end{cases}
\end{equation}
which is $C^\infty$, with all derivatives equal to zero at $ 0  $ and $ T $. The two windows are shown in figure \ref{fig:windows}. These windows'  spectra  exhibit algebraic and non-algebraic decay rates for large frequencies, respectively. 
Their spectral content and an illustration of aliasing effects on them are shown in figure \ref{fig:winspec}. 

When used to estimate signal spectra, for instance when performing a frequency-domain analysis of a system,  the beam-width and dynamic-range of the window are key parameters, as discussed in \cite{harris1978use}.
Higher-order windows tend to be more compact, and thus make a poorer usage of window data, leading to a lower frequency resolution, which is a  trade-off with the improved  convergence rate.

In a periodogram approach, the penalty of this trade-off can be alleviated by window overlap. The typical motivation for window overlap is to increase the number of samples for averaging, or increase the sample length\hide{, as illustrated on figure \ref{fig:effectivewindow}}. This approach comes with the drawback of creating an artificial correlation between samples. It is important to estimate this correlation: if response samples are used to estimate spectral properties, excessive overlap leads to an increase in computational cost without improving the results. 

Sample correlation can be estimated assuming a Gaussian process and a flat spectral content,  and the power spectrum standard variation can be estimated as \cite{welch1967use}
\begin{equation}\label{key}
	\frac{\text{Var}\{ \hat{x}^2 \}} {\text{E}^2\{\hat{x}\}} = \frac{\left( 1+2\sum_{j=1}^{K-1} \frac{K-j}{K}\rho_j \right)}{K}
	\approx  \frac{\left( 1+2\sum_{j=1}^{K-1} \rho_j \right)}{K},
\end{equation}
where 
\begin{equation}\label{key}
	\rho_j = \left(\frac{\int w(t) w(t-jT(1-\tau) ) dt}{\int w^2(t) dt}\right)^2,
\end{equation}
$\tau$ is the window overlap fraction,$ K \approx \frac{L}{T(1-\tau)} $ is the total number of samples,  $ L $ the length of available data and $ T $ the window length. The approximation corresponds to the limit where $ K \gg 1/ \tau $, implying $ \rho_j=0 $ for $ j\gg1 $. 

\begin{figure*}[h]
	\centering
	\includegraphics[width=.9\linewidth,trim={80 0 80 0},clip]{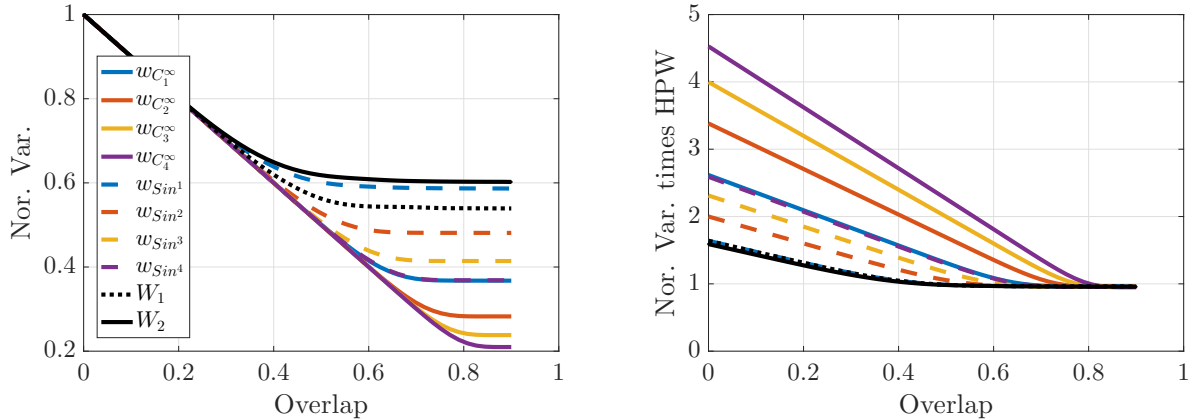}
	\caption{Variance reduction due to overlap. On the left variance is normalized by the zero overlap value (Nor. Var.), on the left this normalized value is multiplied by each window half-power width (HPW). }
	\label{fig:welch}
\end{figure*}

Detailed relations between correlation and window overlap, for a broad class of windows, is available in the literature \cite{harris1978use}. Figure \ref{fig:welch} shows the reduction in standard variation, for a given $ L $, when overlap is used for the windows here studies and for $W_n(t) = 1-(t-0.5)^n $ for $ 0<t<1 $, for reference. Higher-order windows require larger overlaps  for the variance to converge to its minimum value, which is related to their lesser use of window data. Multiplying the variance by the window's half-power width, a measure of the variance in terms of an effective window size is obtained. In terms of this metric, all windows approximately converge to the same variance. For the proposed windows with $ n \le 4 $, a $ 80\% $ overlap guarantees good convergence on the estimation variance. 

In section \ref{sec:sigwindowing} we derive correction terms for signal windowing and explain their use in system identification, with  windowing functions to be used in such framework in order to minimise aliasing proposed in section \ref{sec:window}. In the next subsection we will test the present system identification method with the window families.

\section{Numerical experiments} \label{sec:numTests}
	The proposed method is compared against the approach described in \cite{pintelon1997identification}. A first order system with $n_x=n_u=5$, $ n_a=1$ and $n_b=0$ is used. Assuming $A_0=I$ the system reads,
	\begin{equation} \label{eq:numsys}
	\dfrac{dx}{dt}(t) + A_1 x(t) = B_0 u(t), 
	\end{equation}
	\begin{equation} 
	u(t) = \sum_{j=1}^{n_f} a_j \e^{2\pi i f_j t}.
	\end{equation}
	A total of $n_f=85$ forcing terms with frequencies uniformly spaced between 1 and $20\sqrt{2}$ were used. 
	The elements of the matrices $A_1$, $B_0$ and the force coefficients $a_n$ are taken from a random number generator, as is the initial condition $x(0)=x_0$.  \eqref{eq:numsys} is integrated numerically using a fourth-order Runge-Kunta method and a time step of $3.5 \times 10^{-5}$. These parameters guarantee a very accurate solution, which can be used to evaluate the performance of the approaches.
	
	In total, the model contains 50 parameters to be estimated. In the following subsections, the error in \eqref{eq:sysDiffEq_w2} and the accuracy of the parameter identification on a noiseless system is studied, and later parameter identification on a noisy system is performed. The proposed approach, where the spurious terms are computed, is compared to the approach where a rectangular window is used and these terms are estimated, as described in \cite{pintelon1997identification}. The latter will be referred to as P\&S. To compare the methods on the same basis, a least-square errors procedure is used on both approaches. Note that P\&S with $n_p=0$ corresponds to the scenario where the spurious inputs are ignored.
	
\subsection{Noiseless system}  \label{sec:numTests_noiseless}
\begin{figure*}
	\centering
	\subfloat[]{\includegraphics[width=.5\linewidth]{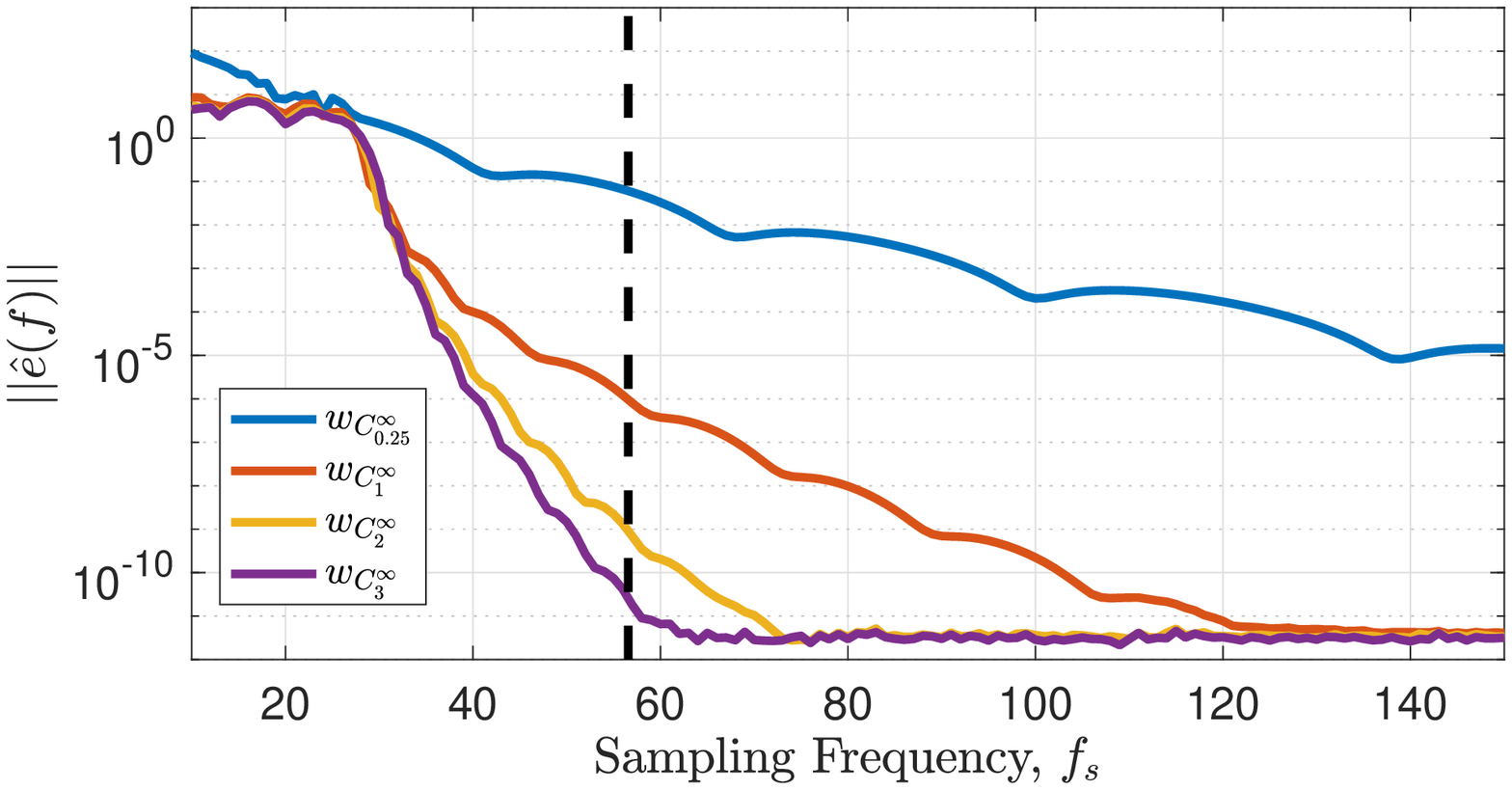} }
	\subfloat[]{\includegraphics[width=.5\linewidth]{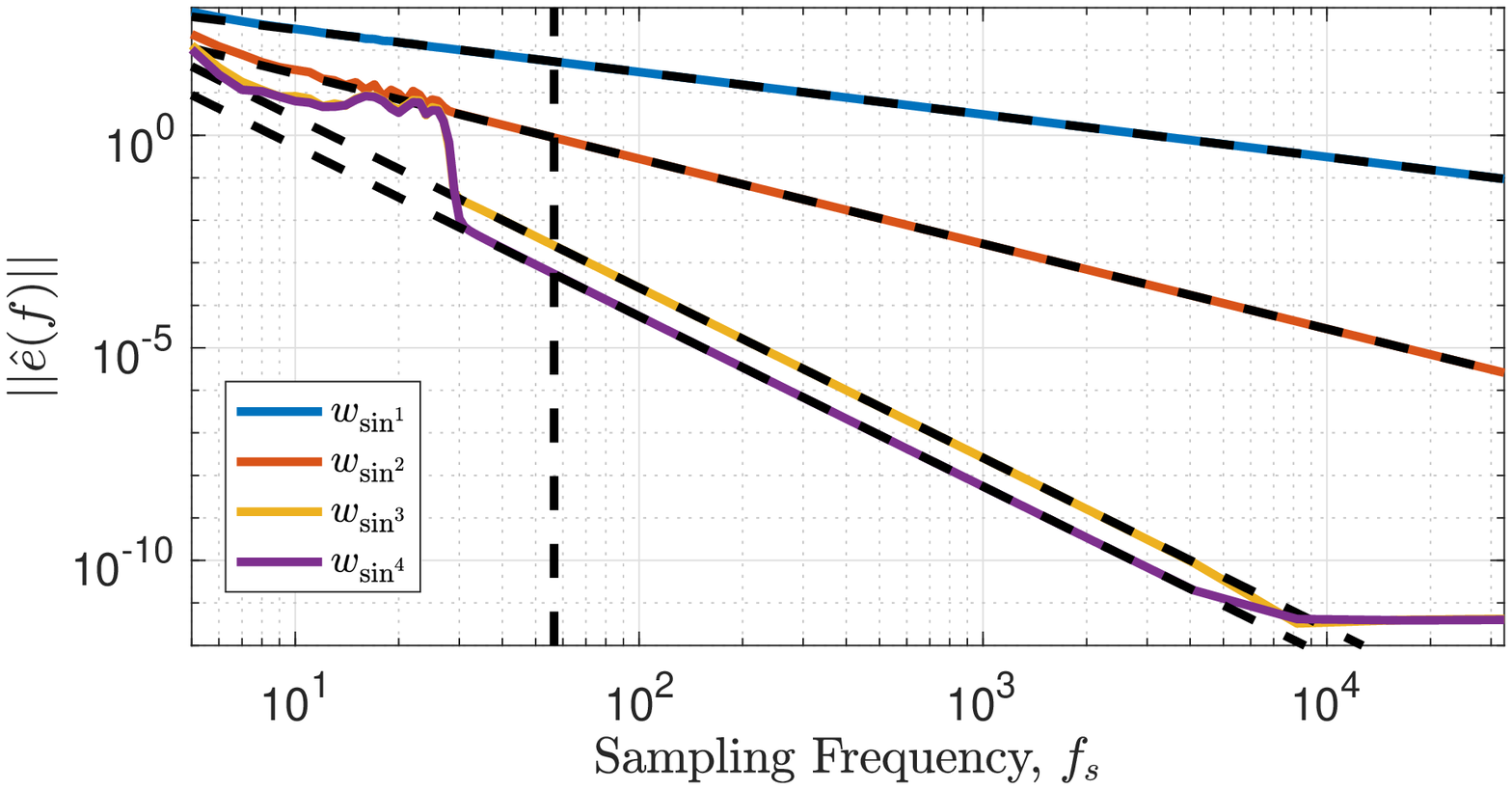} }

	\subfloat[]{\includegraphics[width=.5\linewidth]{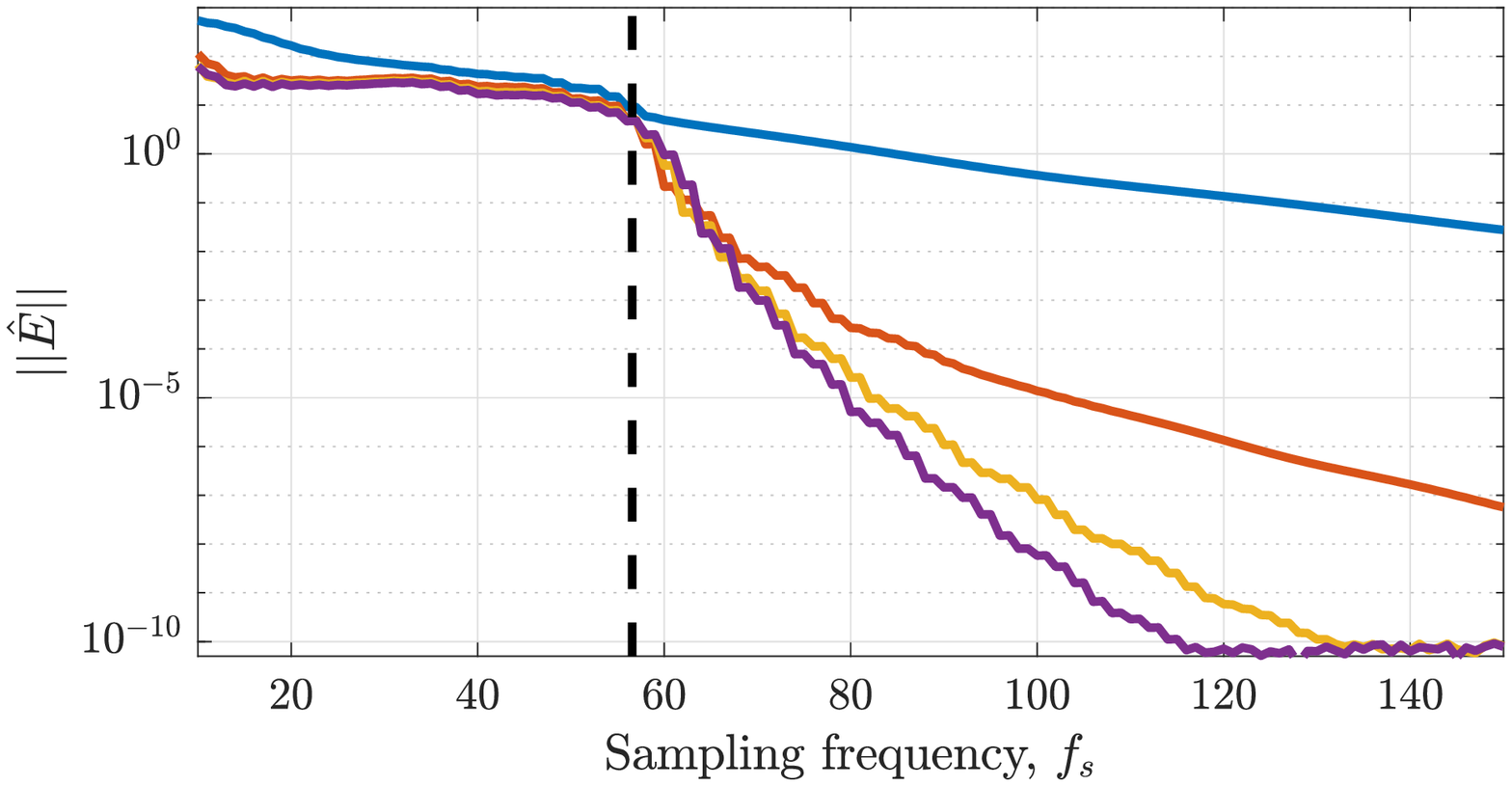} }
	\subfloat[]{\includegraphics[width=.5\linewidth]{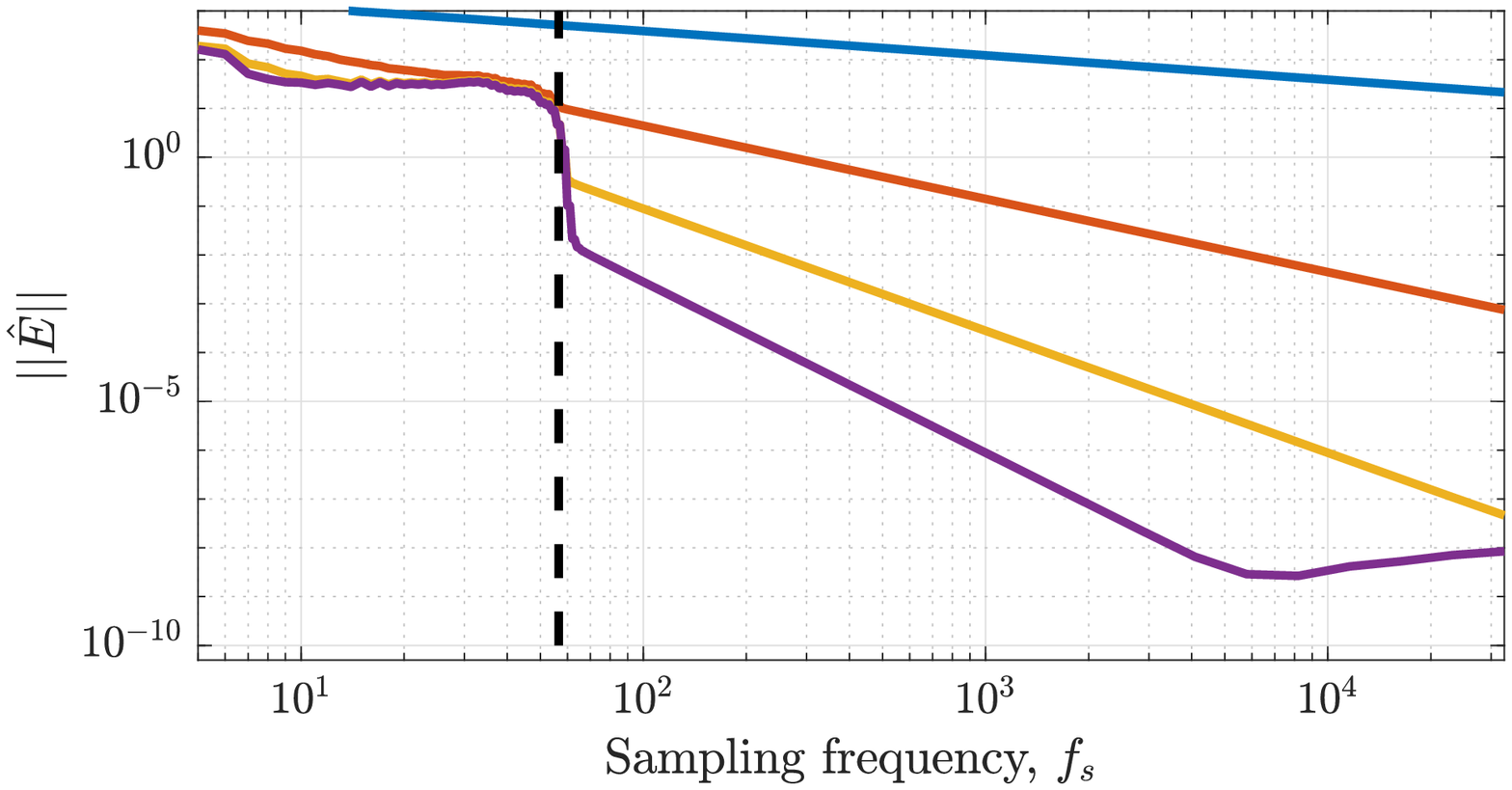} }

	\subfloat[]{\includegraphics[width=.5\linewidth]{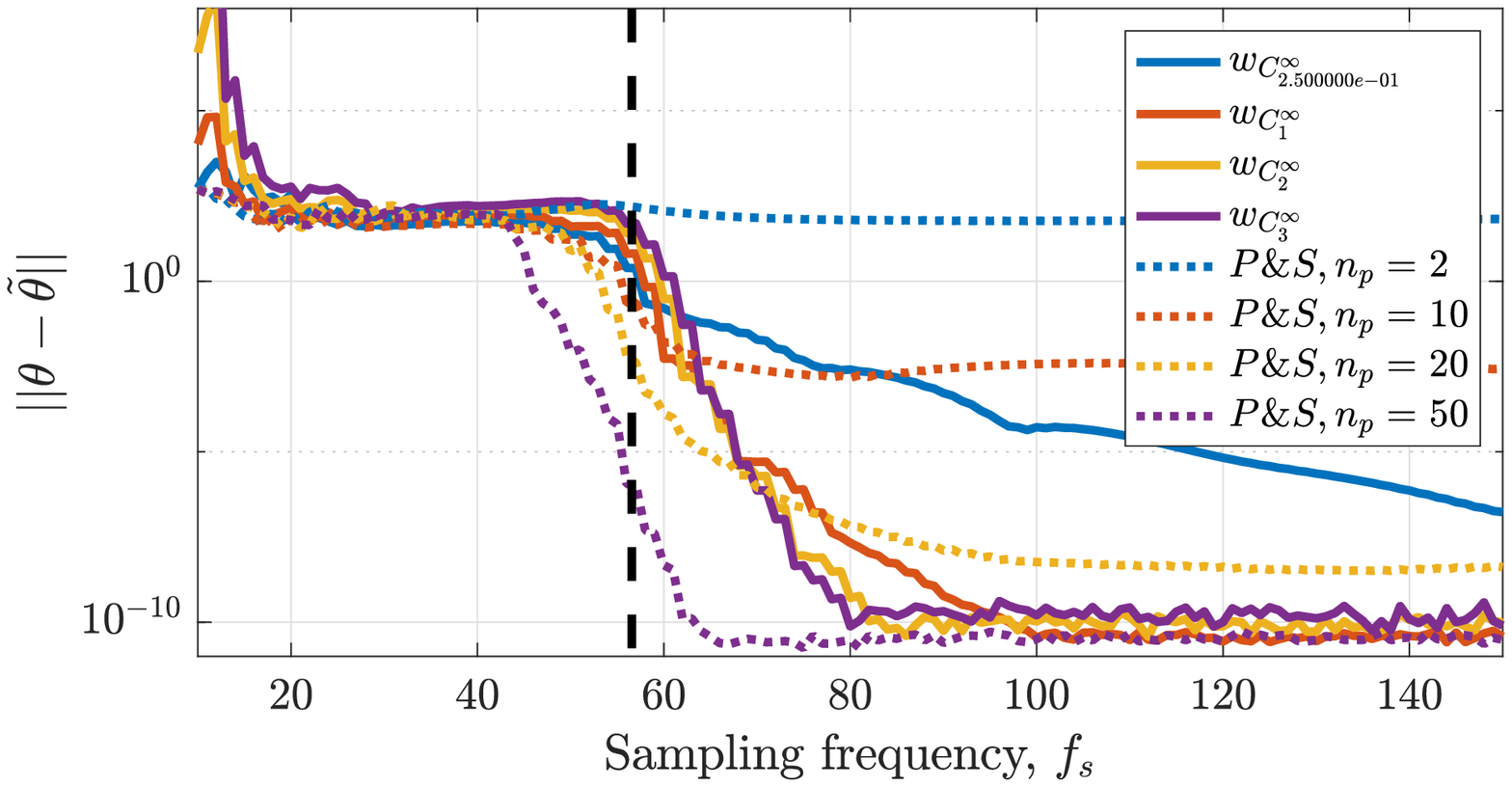} }
	\subfloat[]{\includegraphics[width=.5\linewidth]{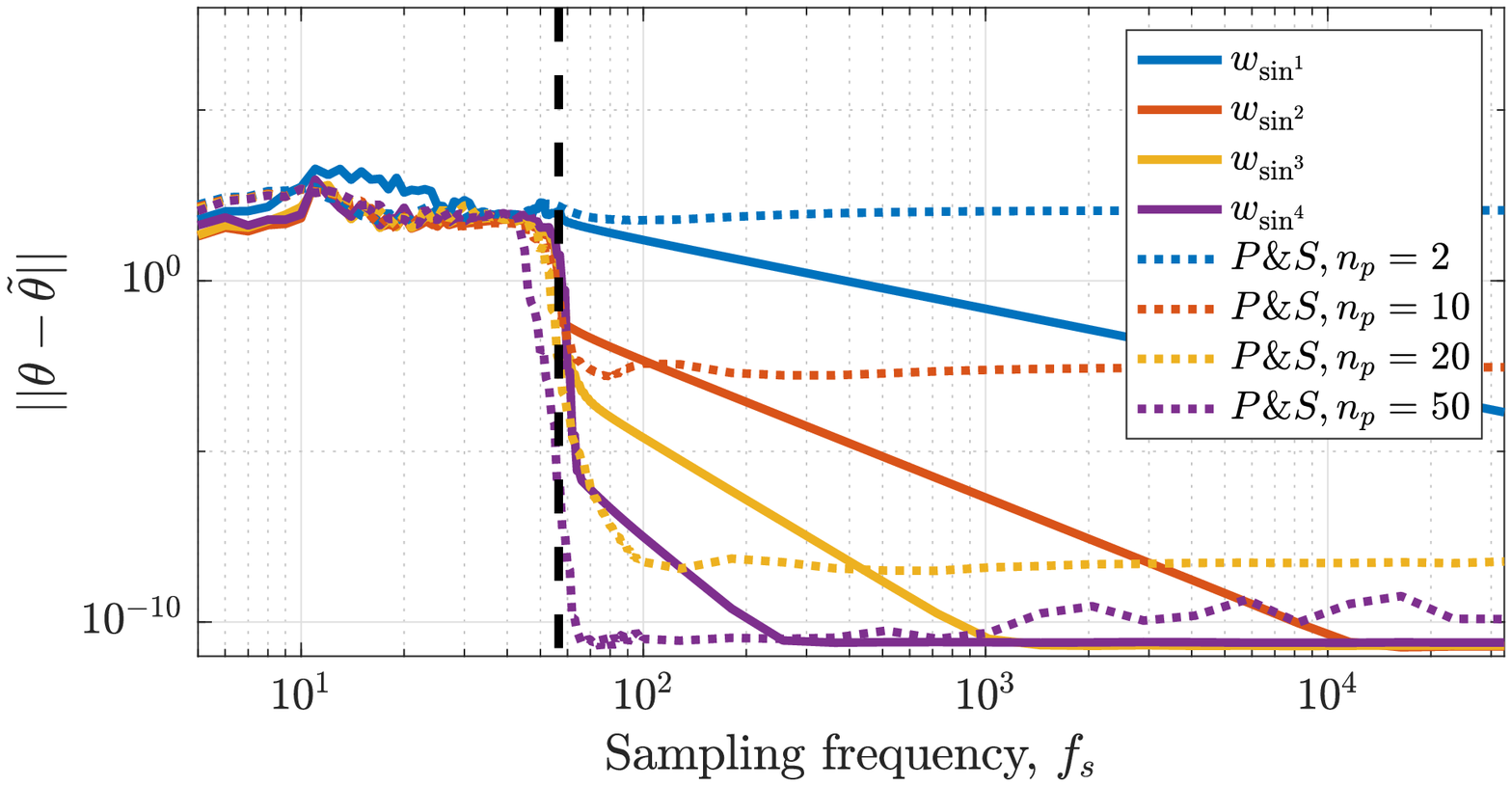} }
	\caption{Errors in \eqref{eq:sysDiffEq_w2}. Error norms at $f=2$ are show in (a,b), the $L_2$ error norms are show in (c,d), and parameter estimation errors in (e,f). The dashed lines in (b) correspond to $1/f_s$ ($w_{\sin^1}$), $1/f_s^2$ ($w_{\sin^2}$), $1/f_s^4$ ($w_{\sin^3}$ and $w_{\sin^4}$)  trends.  Results for the  P\&S approach are shown in (c)-(f) with dotted lines. The vertical dashed line is located at $f_s=40\sqrt{2}$, i.e. $f_s$ for which Nyquist frequency is equal to the maximum excitation frequency.}
	\label{fig:noiselessEst}
\end{figure*}

Figure \ref{fig:noiselessEst} shows the frequency error norm of the system given by \eqref{eq:numsys} as well as the error of the estimated parameters.  The norms are computed as
\begin{eqnarray}
	||\hat e(f)|| &= &\sqrt{  \sum_i |\hat e_{i}(f)|^2 }, \\
	||\hat{ E} || &=& \sqrt{ \int ||\hat e(f)||^2  d f }, \\
	||\theta - \tilde \theta || &=& \sqrt{ \sum_i  \sum_j |\theta_{i,j} - \tilde \theta_{i,j}|^2},
\end{eqnarray}
where $||\hat e(f)|| $  represents  the error norm in \eqref{eq:sysDiffEqfreq} for frequency $f$, $||E||$ the $L_2$ norm of the error, and $ ||\theta - \tilde \theta || $ the error in the estimated system parameters.

\begin{figure}
	\centering
	\includegraphics[width=\linewidth]{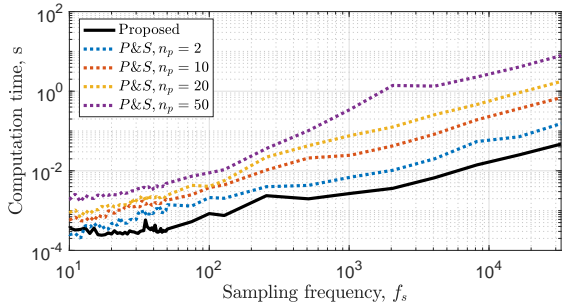} 
	\caption{Computational time of estimation using the proposed and P\&S approaches for different sampling rates.  }
\label{fig:costComparisons}
\end{figure}

\begin{figure}
	\centering
	\includegraphics[width=\linewidth]{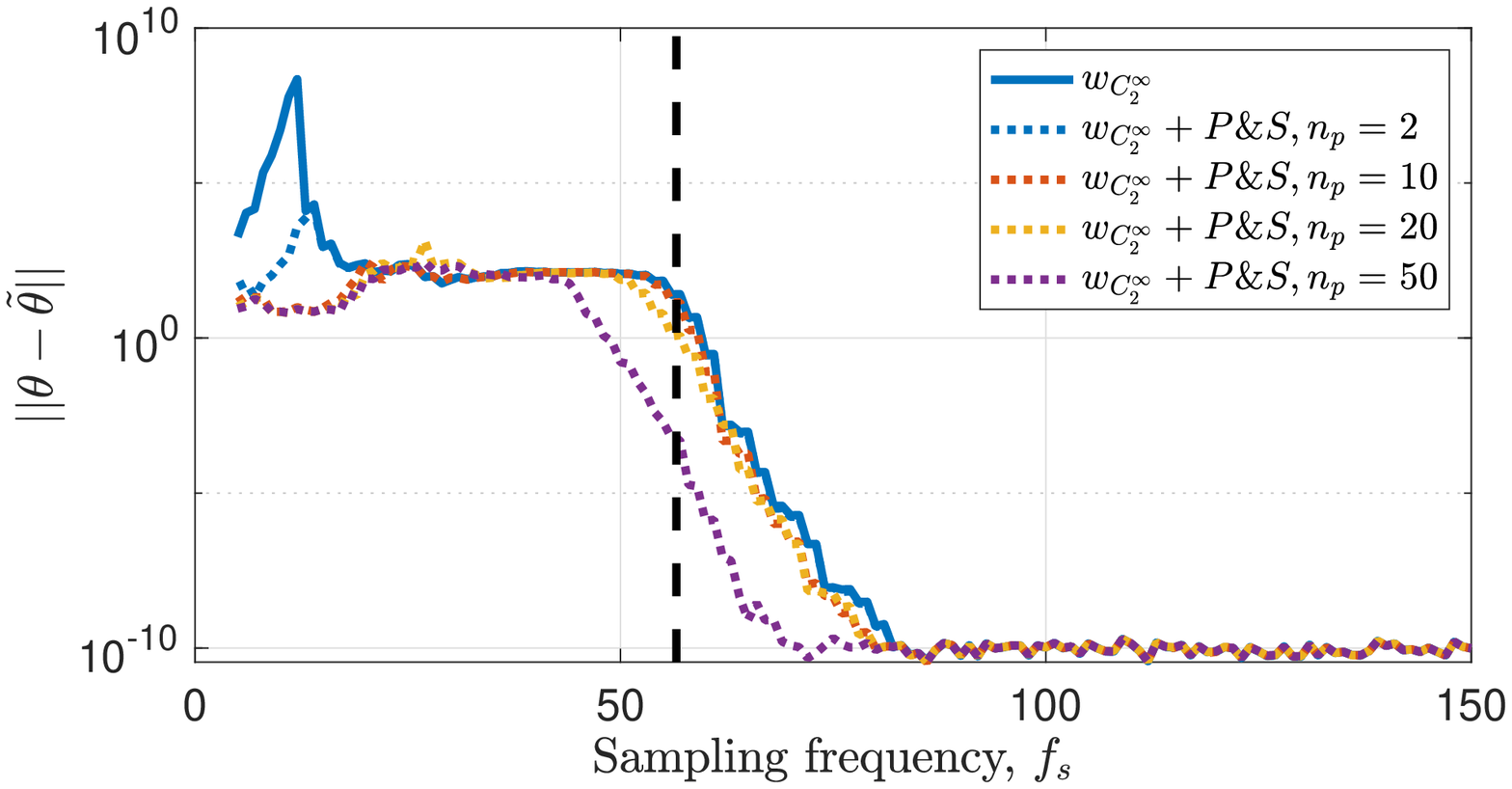} 
	\caption{Same as figure \ref{fig:noiselessEst} (c), using a mixed approach.}
	\label{fig:noiselessEst_mixed}
\end{figure}

Errors in \eqref{eq:sysDiffEqfreq} and in parameter estimations exhibit an algebraic/non-algebraic decay when $w_{\sin^n} / w_{C^\infty_n}$ windows are used, as expected from the window properties discussed in section \ref{sec:window}. Using the infinitely smooth windows, numerical precision is  obtained if the Nyquist frequency is slightly above the maximal signal excitation. The same accuracy is only achieved if a polynomial of order 50 is used in the P\&S approach, in which the estimation of a total of 250 extra parameters is required.  Figure \ref{fig:costComparisons} shows the computational time required by each method, where it is seen that the estimation of the extra parameters considerably increases the total costs. A mixed approach, where polynomials terms are estimated to reduce the impact of aliasing effects on the estimation, does not provide additional gains over the individual approaches, as  shown in figure \ref{fig:noiselessEst_mixed}.

As described in section \ref{sec:aliasing},the errors decay with the smoothness of the windowed signal. If the window is a function $C^n$, and $max(n_a,n_b)=1$,  $||\hat{E}||$ and $||\theta-\tilde{\theta}||$ show a decay with $1/f_s^{n}$.  The expected decay rate of $1/f_s^{n+1}$ for $||\hat{e}(f)||$ is observed for $w_{sin^3}$ but not for $w_{sin^1}$.  This is due to the fact that $(dw/dt x)(0) \neq (dw/dt x)(T)$, and thus the FFT transform is not equivalent to the trapezoidal rule, and the results of \cite{trefethen2014exponentially} are not applicable. This can be remedied by averaging out the signal values at the begging and end of the window.


\subsection{Estimation in a noisy system}  \label{sec:numTests_noisy}

\begin{figure*}
	\centering
		\subfloat[$\sigma=10^{-2}$]{\includegraphics[width=.5\linewidth]{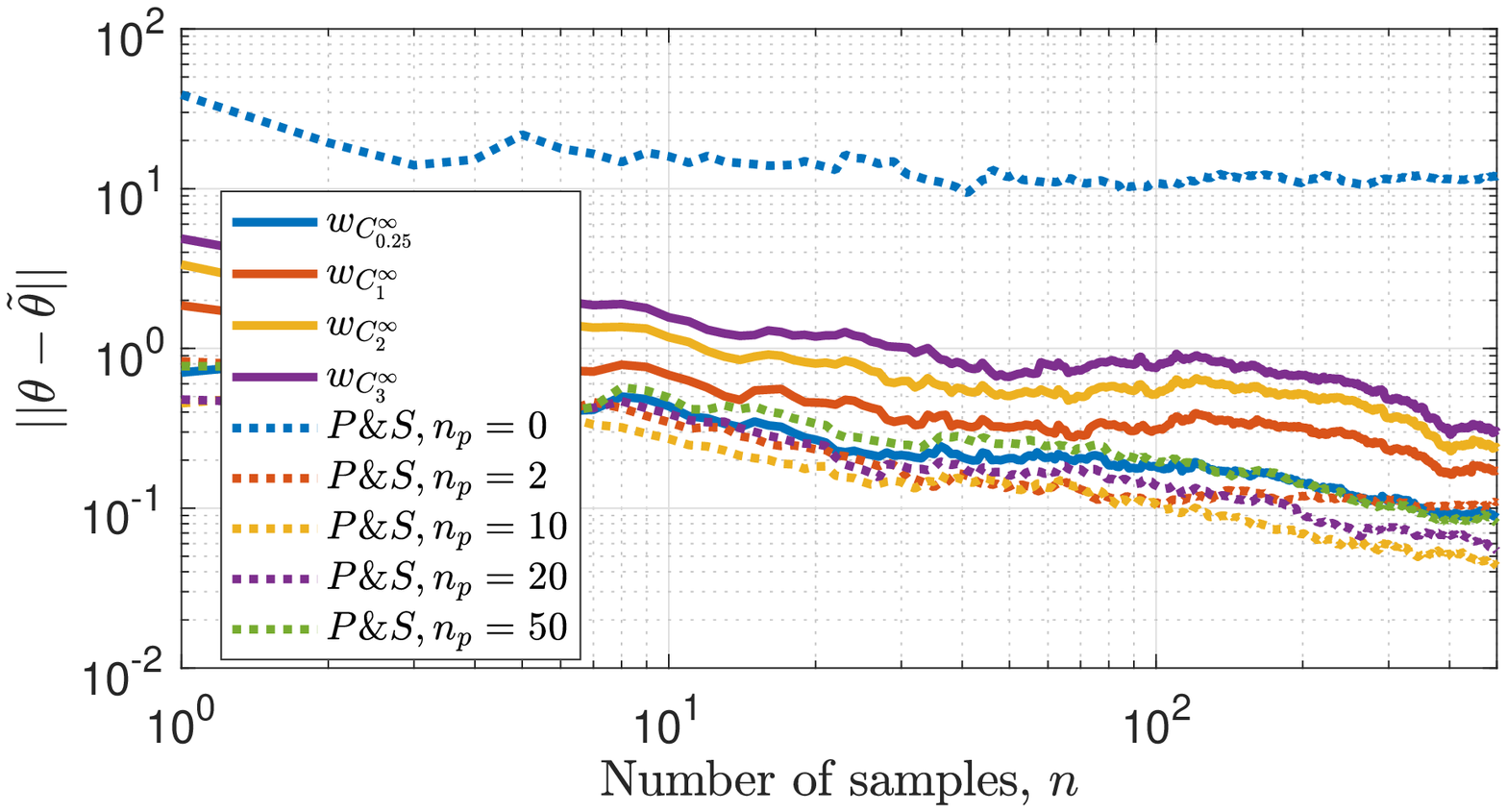}}
		\subfloat[$\sigma=10^{-8}$]{\includegraphics[width=.5\linewidth]{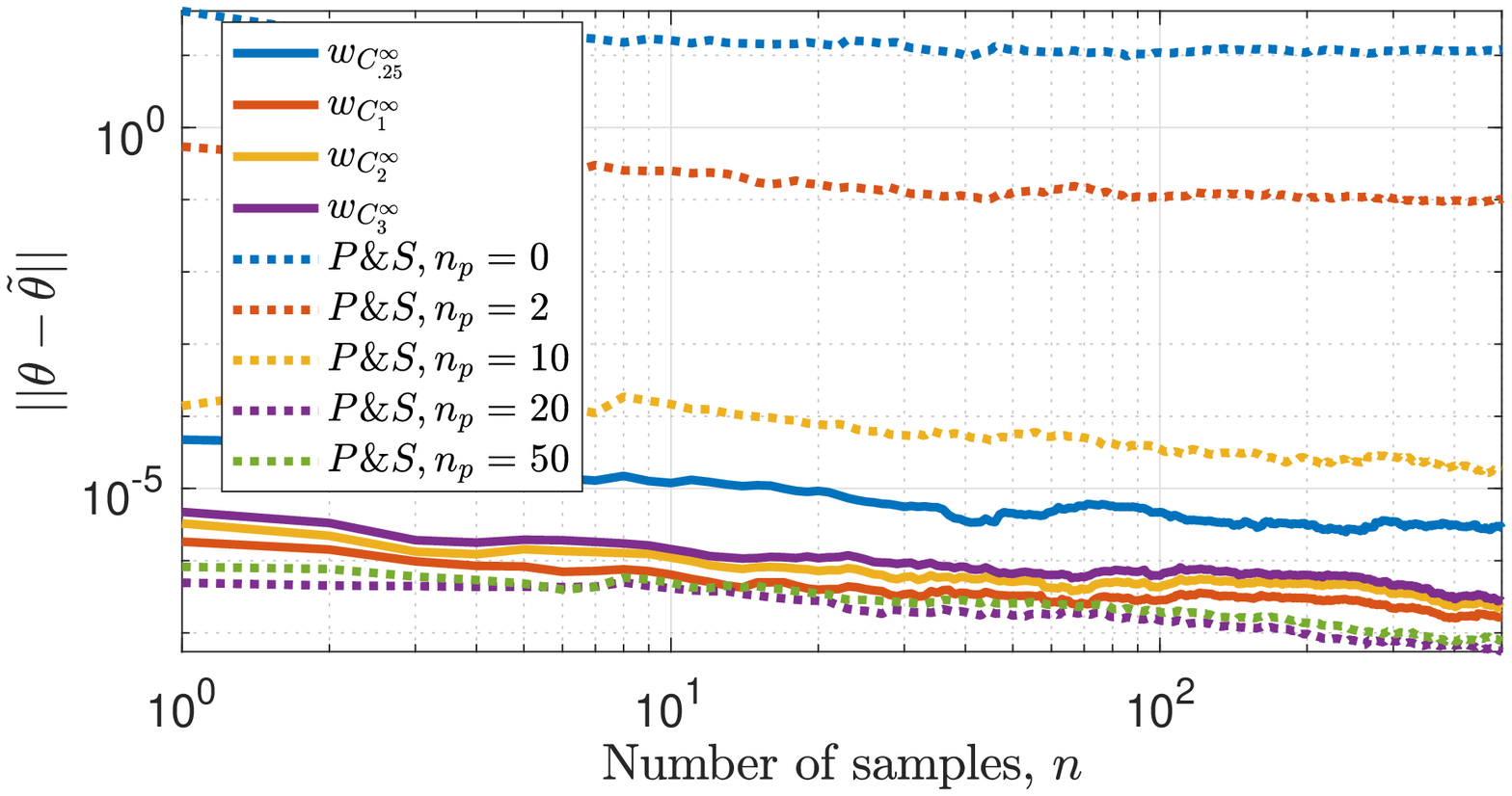} }

		\subfloat[$\sigma=10^{-2}$]{\includegraphics[width=.5\linewidth]{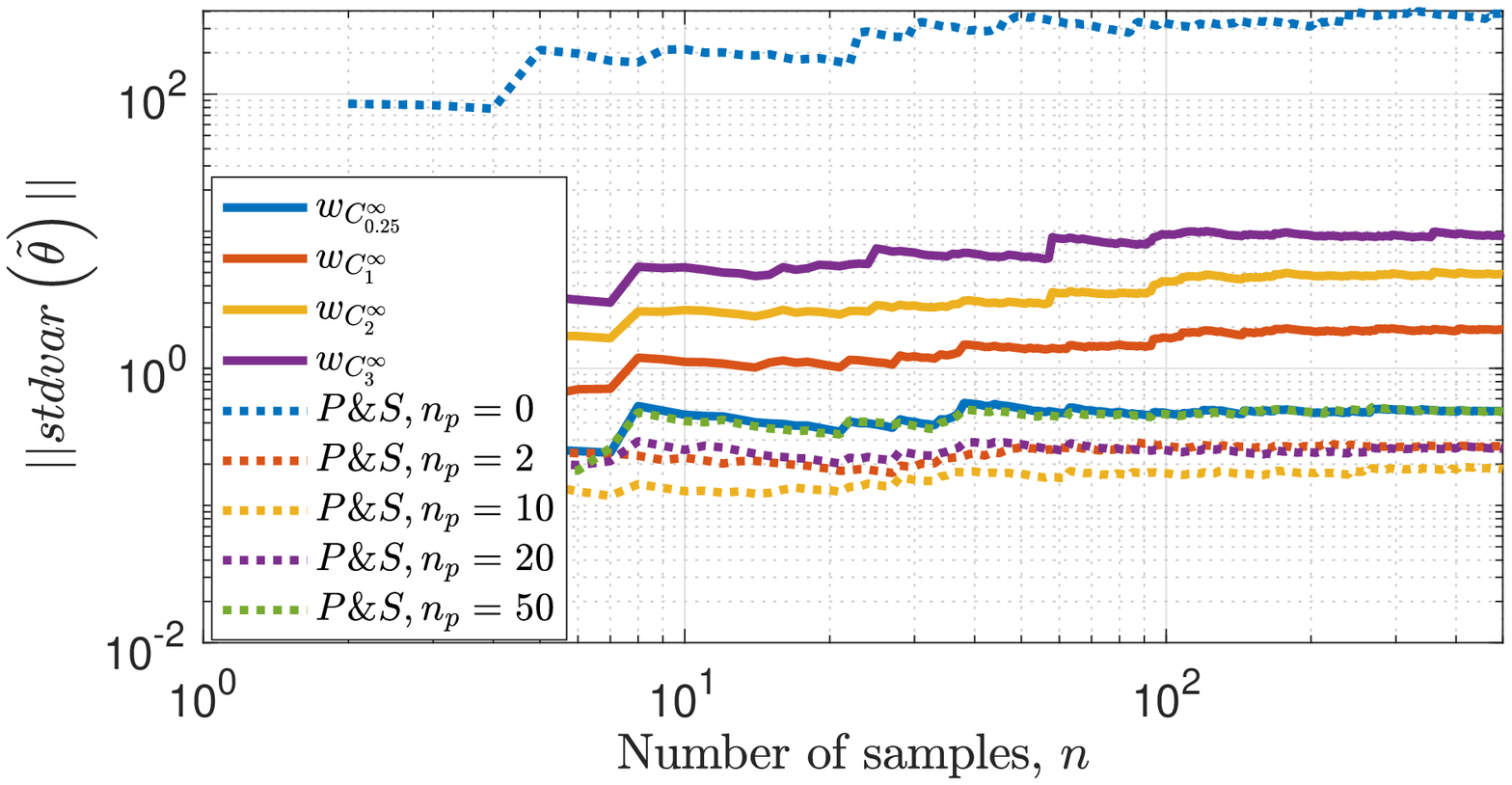} }
		\subfloat[$\sigma=10^{-8}$]{\includegraphics[width=.5\linewidth]{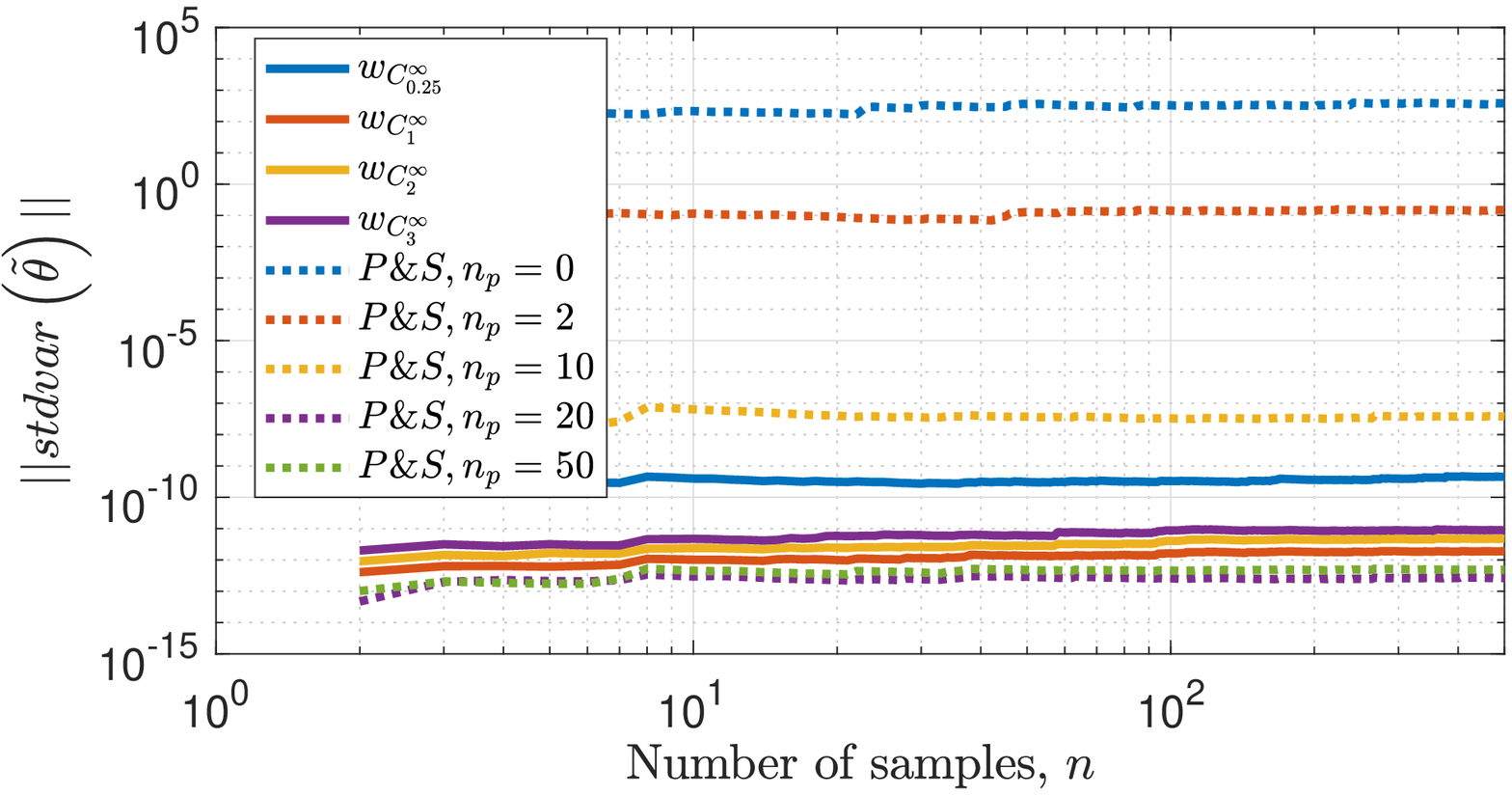} }

	\caption{Parameter estimation standard deviation and error as a function of the number of samples for a sampling frequency of $f_s=80$.}
	\label{fig:noisyEst1}
\end{figure*}

We consider now estimations on a noisy environment. A total of 500 data sets were used, with signals $x$ and $u$  corrupted with white noise with standard variation  $\sigma$.  Parameters were estimated  for each sample. The error between the mean value of the parameters and their standard variations are shown in figure \ref{fig:noisyEst1} as a function of the number of samples used.

The performances of the different windows in the high noise scenario ($\sigma=10^{-2}$) and in the  lower noise scenario ($\sigma=10^{-8}$) are inverted, with lower values of the parameter  $n$ leading to more accurate estimation on the former, and higher values on the latter. This trend is associated with the better use of the available data by the window for low values of $n$ (see figure \ref{fig:windows}), being thus able to better account for noise, and with the smaller aliasing effects on higher order windows (see appendix \ref{app:aliasing}).  The trade-off between lower type II aliasing and noise errors depends on the signal spectral content and the noise levels. The optimal choice is thus problem dependent. Note that although the least-squares method used here is biased, the reduction of the estimation errors with the number of sample size in figure \ref{fig:noisyEst1}(a) indicates that the variance of the accuracy of the estimation is larger than the bias. Such bias can be removed using maximum-likelihood estimations, e.g.  as used in  \cite{pintelon1997identification}.

\section{Conclusion}\label{sec:conclusion}
A new interpretation of windowing errors in frequency domain representation of ODEs has been proposed, together with a correction technique applicable to arbitrary window functions. Two types of windows were explored, each leading to an algebraic and a non-algebraic decay of errors associated with aliasing effects when sampling frequency is increased. 

The presented work can be used in a frequency-domain investigation of systems, e.g. as in \cite{nogueira2020forcing}, where correspondence between the system's inputs and outputs via the linear operator is fundamental for the investigation of the relevant physical mechanisms, or for purposes of system identification. For low-noise systems, the method exhibits better performance and/or lower costs than the P\&S approach, proposed in \cite{pintelon1997frequency,pintelon1997identification}. 

In the proposed approach, signal windowing leads to noise at different frequencies to be correlated. Although the construction of a maximum-likelihood estimators in this case is considerably more complicated than the one proposed by  \cite{pintelon1997frequency,pintelon1997identification}, several methods are available in the literature to obtain an unbiased estimation in these cases, e.g. \cite{soderstrom2011generalized,soderstrom2017errorsinvariables}.  In the current work we show that with the exact representation of the system obtained, a simple least-squares estimate is seen to provide accurate and cheap parameter estimates when the system is noise-free, or when noise levels are small.
Also, \eqref{eq:sysDiffEq_w2} can be used to extend methods originally designed for periodic signals, e.g. \cite{mi2012frequencydomain}, be used with arbitrary signals.
An extension to systems of partial differential equations can be constructed using external products of the proposed windows, such as $ w_{2D}(x,y)=w_{C_n^\infty}(x)w_{C_n^\infty}(y) $, as in \cite{asiri2017source}. 

The novel infinity-smooth window results in aliasing effects orders of magnitude lower than classical windows,  as shown in appendix \ref{app:aliasing},  leading to considerably more accurate identification requiring only moderate sampling rates, being thus a quasi-optimal window choice for such an application. For noisy systems, it was observed that a  trade-off between a better use of the available data, i.e. lower values of $n$, and lower aliasing-effects, i.e. higher values of $n$, depends on the magnitude of these factors, and is thus problem dependent, as illustrated in section \ref{sec:numTests_noisy}. 

\section*{Funding}
E. Martini acknowledges financial support by CAPES grant 88881.190271/2018-01.
Andr\'e V. G. Cavalieri was supported by CNPq grant 310523/2017-6. 

\appendix
\section{Aliasing effects} \label{app:aliasing}

\begin{figure*}[t!]
	\centering
	\includegraphics[width=\linewidth,trim={0 8px 0 8px},clip]{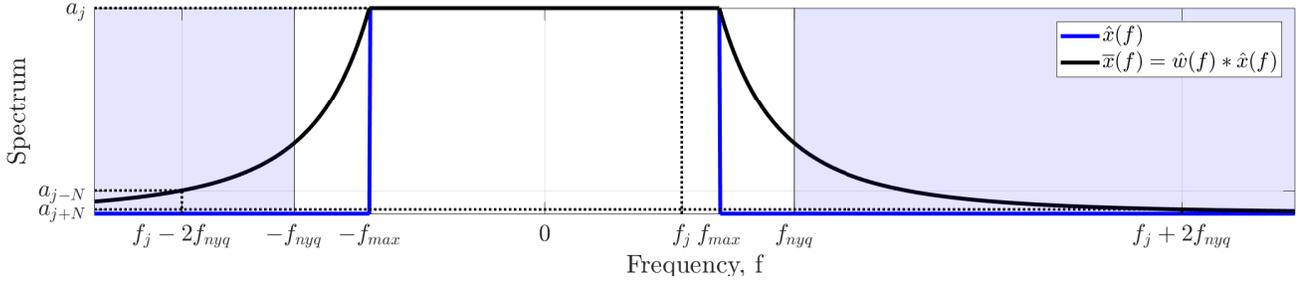}
	\caption{illustration of aliasing effects of windowed signals.  A window function $ w $ is applied to a band-limited signal (blue line representing its frequency content), resulting in spectral leakage, which spreads the frequency content of the signal (shown in black). The signal is sampled with Nyquist frequency $ f_{nyq}=N/2T $; blue region indicates unresolved frequencies. The frequency content at $ f_j=j/T $, $ a_{j} $, and its aliased components $ a_{j+N},a_{j-N} $ are indicated.}
	\label{fig:windowleakage_alias}
\end{figure*}

An analytical expression for the aliasing effects on the Fourier transform of a windowed signal  is derived.  Most of these results are a direct consequence of the results presented in \cite{trefethen2014exponentially}, which are summarized below. For simplicity we assume $s=(wx)(t)$, with results for $u(t)$ and derivatives of $w(t)$ being analogous.

The Fourier-series representation of the windowed signal reads,
\begin{equation}\label{eq:fourierseries}
s(t) = \sum_{k=-\infty}^\infty a_k \mathrm{e}^{{2\pi \mathrm{i} k} t /T },
\end{equation}
with
\begin{equation}\label{eq:a_k}
a_k=\dfrac{1}{T}\int_0^T s(t) \mathrm{e}^{-{2\pi \mathrm{i} k} t /T} dt,
\end{equation}
or equivalently, 
\begin{equation}\label{eq:a_k_conv}
a_k = \int_{-\infty}^\infty \hat w(f)\hat x \left(\frac{k}{T}-f\right) df.
\end{equation}
From this equation the contribution of windowing to aliasing effects, i.e. type II aliasing, as defined in section \ref{sec:aliasing}, is seen explicitly. This effect is illustrated in figure \ref{fig:windowleakage_alias}.

The coefficients $ \tilde a_{k,N} $ of an  $ N $-point discrete Fourier transform are related to $a_k$ via, 
\begin{equation}\label{eq:a_nk}
\begin{aligned}
\tilde a_{k,N} &= \dfrac{1}{N}\sum_{j=1}^N  s\left(\frac{jT}{N}\right) \mathrm{e}^{-2\pi\ii k j /N }
\\& =   \sum_{m=-\infty}^\infty a_m \dfrac{1}{N}\sum_{j=1}^N 
\mathrm{e}^{-2\pi \ii  (m+k) j /N} 
\\& = a_k + \sum_{m=1}^{\infty} \left( a_{k+mN} + a_{k-mN}\right),
\end{aligned}
\end{equation}              
where the sum in the final line represents the aliasing effects arising from unresolved frequencies (components for which $ |k|>N/2 $), i.e. the blue region in figure \ref{fig:windowleakage_alias}.

For a band-limited signal, \eqref{eq:a_nk} shows the aliasing effects on the term $ \tilde{a}_{k,N}$.  Using windows for which $ \hat w(f) $ decays faster will lead to smaller aliasing effects, and thus to faster convergence of $ \tilde a_{k,N} $ to  $a_k$.  As $a_k = O(1/k^n)$ for $s(t)\in C^n$, then $|a_k-\tilde{a}_{k,N}| = O(1/k^n)$.  

A slightly stronger results holds if $s$ is smooth between 0 and T. For this case $|a_k-\tilde{a}_{k,N}| = O(1/k^{n+1})$ for off $n$ \cite{trefethen2014exponentially}, which is due to a partial cancellation between $ a_{k+mN} + a_{k-mN} $ in \eqref{eq:a_nk}.

\hide{\color{red}
The window functions considered in this study are illustrated in figure \ref{fig:windows}, and they have been shown in section \ref{sec:FreqDomainIdErrors} to yield polynomial and exponential convergence rates. The $ w_{\sin^n} $
window family has in fact been widely used in the literature: $ w_{\sin^1} $ corresponds to the sine
(sometimes refereed to as cosine) window; $ w_{\sin^2} $ is the Hann
window; and $ w_{\sin^n} $ has been used as the $ \cos^\alpha $ window in Ref. [13]. The fast spectral
decay for large frequencies has already been described \cite{harris1978use}, and it has been remarked that spectral leakage is
minimized by setting window values and their derivatives to
zero at the window extremes. However, to the best of our
knowledge, this approach has never been applied to frequency domain or modulating function identification. Recent
frequency domain studies employ rectangular windows \cite{goos2017frequency},
and modulating functions are usually taken to have vanishing derivatives only up to the order of the system equation \cite{asiri2017modulating}. We are unaware of studies that employ window functions with zero derivatives at all orders, such as $ w_{C^\infty_n} $.
}

\subsection{Estimate of aliasing effects }
An estimation of the aliasing effects can be obtained by considering a band-limited signal with $ \hat x(|f|>f_{max})=0 $.  The coefficient of the leading term, is  $\max(|a_{k+N}|, |a_{k-N}|)$,  where
\begin{align}\label{}
a_{k\pm N} & = \int_{-f_{max}}^{f_{max}} \hat w(k\pm N-f) \hat x (f) df
\end{align}

\begin{figure}[t]
	\centering
	\includegraphics[width=\linewidth,trim={0 2px 0 11px },clip]{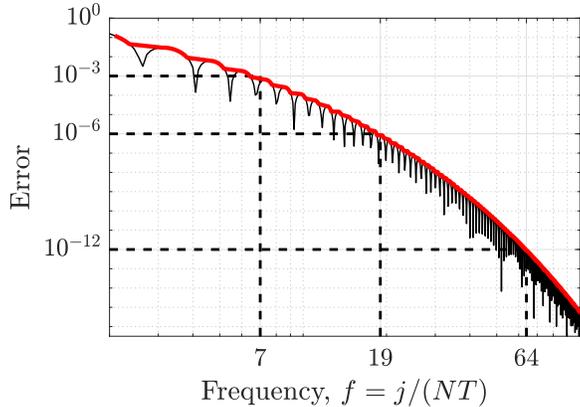}
	\caption{Illustration of the determination of $ f^{err}_p  $ for the $ w_{C^\infty_1} $ window. Black lines correspond to $ \hat w(f)/S $, where $ S $ is the window area. Red line indicates the error envelope and dashed error levels and their corresponding $ f_p^{err} $.}
	\label{fig:ferrc2}
\end{figure}

We define $ f_p^{err} $ as the smallest frequency for which $ |w(|f|>f_p^{err})|/S <p $, where $ S $ is the area under the window. Thus, by choosing $ N $ such that $ |k\pm (N-f_{max})|>f_p^{err} $, the influence of each frequency component of $ \hat x(f) $ on $ a_{k\pm N} $ is smaller than $p$. The process is illustrated in figure \ref{fig:ferrc2}. Derivation for the window derivatives is analogous. and   $ f_p^{err} $  values for the proposed windows and their derivatives are provided in table~\ref{tab:WindowsFerr}. 

\begin{table*}
	\centering
	\caption{Values of $ f^{err}_p $ for the proposed windows. }
	\resizebox{\textwidth}{!}{
		\begin{tabular}{l||r|r|r||r|r|r||r|r|r||r|r|r|} \label{tab:WindowsFerr}
			& \multicolumn{3}{c||}{ Window } & \multicolumn{3}{c||}{ Window Derivative } & \multicolumn{3}{c||}{ Window 2nd Derivative } & \multicolumn{3}{c|}{ Window 3rd Derivative }\\
			&  $ f^{err}_{10^{-3}} $   & $f^{err}_{10^{-6}} $ & $ f^{err}_{10^{-12}} $ &  $ f^{err}_{10^{-3}} $   & $f^{err}_{10^{-6}} $ & $ f^{err}_{10^{-12}} $ &  $ f^{err}_{10^{-3}} $   & $f^{err}_{10^{-6}} $ & $ f^{err}_{10^{-12}} $&  $ f^{err}_{10^{-3}} $   & $f^{err}_{10^{-6}} $ & $ f^{err}_{10^{-12}} $ \\ 
			\hline 
			$ w_{\sin^1} $  &  16  & 502  & $>$10000  & 1637  & $>$10000  & $>$10000  &  -  & -  & -   & -  & -  & - \\
			$ w_{\sin^2} $  &   7  &  68  & 4911  &  45  & 1453  & $>$10000  & $>$10000  & $>$10000  & $>$10000  & -  & -  & - \\
			$ w_{\sin^3} $  &   5  &  28  & 867  &  16  & 153  & $>$10000  & 150  & $>$10000  & $>$10000  & $>$10000  & $>$10000  & $>$10000 \\
			$ w_{\sin^4} $  &   4  &  17  & 264  &  10  &  53  & 1683  &  37  & 369  & $>$10000  & 564  & $>$10000  & $>$10000 \\
			$ w_{\sin^5} $  &   5  &  13  & 124  &   8  &  30  & 467  &  20  & 109  & 3491  &  96  & 956  & $>$10000 \\
			$ w_{\sin^7} $  &   5  &  10  &  51  &   7  &  17  & 116  &  12  &  35  & 346  &  26  & 102  & 1607 \\
			\hline\hline
			$ w_{C^\infty_{0.25}}$&  12  &  45  & 191  &  34  &  99  & 320  &  93  & 198  & 507  & 202  & 354  & 760 \\
			$ w_{C^\infty_1} $  &   7  &  19  &  64  &  14  &  33  &  95  &  28  &  55  & 136  &  50  &  88  & 187 \\
			$ w_{C^\infty_2} $  &   7  &  15  &  41  &  11  &  22  &  57  &  19  &  34  &  77  &  30  &  50  & 103 \\
			$ w_{C^\infty_3} $  &   7  &  13  &  33  &  10  &  19  &  44  &  16  &  27  &  58  &  24  &  38  &  76 \\
			$ w_{C^\infty_4} $  &   7  &  13  &  30  &  10  &  18  &  39  &  15  &  24  &  50  &  21  &  33  &  63 \\		
			\hline 
	\end{tabular} }
\end{table*}

\bibliographystyle{plain}        
\bibliography{References}           

\end{document}